\documentclass[sigconf]{acmart}

\AtBeginDocument{%
  }

\copyrightyear{2025} 
\acmYear{2025} 
\setcopyright{cc}
 \setcctype{by}
 \acmConference[ISCA '25]{Proceedings of the 52nd Annual International
 Symposium on Computer Architecture}{June 21--25, 2025}{Tokyo, Japan}
 \acmBooktitle{Proceedings of the 52nd Annual International Symposium on
Computer Architecture (ISCA '25), June 21--25, 2025, Tokyo,
 Japan}\acmDOI{10.1145/3695053.3731002}
 \acmISBN{979-8-4007-1261-6/2025/06}

\usepackage{pifont}
\usepackage{multirow}
\usepackage{enumitem}
\usepackage{xspace}
\usepackage{algorithm}
\usepackage{algorithmic}
\usepackage{listings}
\usepackage{xcolor}
\usepackage{makecell}
\usepackage{color}

\newcommand{\DAHU}[1]{\textcolor{black}{#1}}

\newcommand{\FEH}[1]{\textcolor{black}{#1}}
\newcommand{\feh}[1]{\textcolor{black}{#1}}
\newcommand{\minor}[1]{\textcolor{black}{#1}}
\newcommand{\isca}[1]{\textcolor{black}{#1}}

\newcommand{\rebuttal}[1]{\textcolor{black}{#1}}

\newcommand{\myparagraph}[1]{\vspace {3pt}\noindent\textbf{#1}}
\newcommand{\sys}{vNPU\xspace} %need to include \usepackage{xspace}

\begin{document}

\title{\isca{Topology-Aware Virtualization over Inter-Core Connected Neural Processing Units}}

\author{Dahu Feng}
\authornote{Both authors contributed equally to this research.}
\email{fengdh21@mails.tsinghua.edu.cn}
\affiliation{%
%   \department{Center for Brain-Inspired Computing Research}
  \institution{CBICR, Tsinghua University}
  \city{Beijing}
  \country{China}
}

\author{Erhu Feng}
\email{fengerhu1@sjtu.edu.cn}
\authornotemark[1]
\affiliation{%
    % \department{Institute of Parallel and Distributed Systems}
    \institution{IPADS, Shanghai Jiao Tong University}
    \city{Shanghai}
    \country{China}
}

\author{Dong Du}
\email{Dd_nirvana@sjtu.edu.cn}
\affiliation{%
    % \department{Institute of Parallel and Distributed Systems}
    \institution{IPADS, Shanghai Jiao Tong University}
    \city{Shanghai}
    \country{China}
}

\author{Pinjie Xu}
\email{xupinjie321@outlook.com}
\affiliation{%
    \department{SenseTime Research}
    \city{Beijing}
    \country{China}
}

\author{Yubin Xia}
\email{xiayubin@sjtu.edu.cn}
\affiliation{%
    % \department{Institute of Parallel and Distributed Systems}
    \institution{IPADS, Shanghai Jiao Tong University}
    \city{Shanghai}
    \country{China}
}

\author{Haibo Chen}
\email{haibochen@sjtu.edu.cn}
\affiliation{%
    % \department{Institute of Parallel and Distributed Systems}
    \institution{IPADS, Shanghai Jiao Tong University}
    \city{Shanghai}
    \country{China}
}

\author{Rong Zhao}
\email{r_zhao@tsinghua.edu.cn}
\authornote{Corresponding author.}
\affiliation{%
  \department{Center for Brain-Inspired Computing Research, IDG/McGovern Institute for Brain Research and Department of Precision Instrument}
  \institution{Tsinghua University}
  \city{Beijing}
  \country{China}
}
% \additionalaffiliation{%
%   \institution{Optical Memory National Engineering Research center, IDG/McGovern Institute for Brain Research, and Department of Precision Instrument, Tsinghua University} % 附加单位的机构
% %   \department{Tsinghua University} % 附加单位的部门
% %   % 你可以写 city, country 等，但它们不会在 additionalaffiliation 的注脚中显示
% %   \city{Beijing}
% %   \country{China}
% }

\renewcommand{\shortauthors}{Feng et al.}

% \author{ISCA 2025 Submission \#72 - Confidential Draft - Do NOT Distribute!!}
%\author{...} % removed for anonymity
\begin{abstract}
\isca{With the rapid development of artificial intelligence (AI) applications, 
an emerging class of AI accelerators, termed Inter-core Connected Neural Processing Units (NPU), 
has been adopted in both cloud and edge computing environments, like Graphcore IPU, Tenstorrent, etc. 
Despite their innovative design, these NPUs often demand substantial hardware resources, 
leading to suboptimal resource utilization due to the imbalance of hardware requirements across various tasks.
To address this issue, prior research has explored virtualization techniques for monolithic NPUs,
but has neglected inter-core connected NPUs with the hardware topology.}
% Although previous research has investigated (para-)virtualization techniques to improve the resource utilization for monolithic NPUs, 
% these efforts do not consider critical characteristics of inter-core connected NPUs, such as the interconnection among NPU cores as well as the NPU topology.}

\isca{This paper introduces \sys, the first comprehensive virtualization design for inter-core connected NPUs, integrating three novel techniques: 
(1) NPU route virtualization, which redirects instruction and data flow from virtual NPU cores to physical ones, creating a virtual topology;
(2) \minor{NPU memory virtualization}, designed to minimize translation stalls for SRAM-centric and NoC-equipped NPU cores, 
thereby maximizing the memory bandwidth; 
and (3) Best-effort topology mapping, which determines the optimal mapping from all candidate virtual topologies, 
balancing resource utilization with end-to-end performance.
We have developed a prototype of \sys on both an FPGA platform (Chipyard+FireSim) and a simulator (DCRA).}
\rebuttal{Evaluation results demonstrate that when executing multiple NPU workloads on virtual NPUs, 
\sys achieves performance improvements of up to \DAHU{1.92x} and \DAHU{1.28x} for the Transformer and ResNet models, respectively, 
in comparison to the MIG-based virtualization method. 
Furthermore, the hardware performance overhead associated with the virtualization itself is minimal, 
incurring less than \DAHU{1\%} reduction in end-to-end performance.}
% Evaluation results indicate that, compared to other virtualization approaches such as unified virtual memory and MIG, 
% \sys achieves up to a 2x performance improvement across various ML models, with only 2\% hardware cost.
% \FEH{\sys achieves a upto 2x performance improvement in different ML models}.

\end{abstract}

    

\maketitle % should come after the abstract
\begin{sloppypar}
    % 设置从第二页开始显示页码1
% \setcounter{page}{1} 
% \pagestyle{plain} % 第二页及后续页显示页码

\section{INTRODUCTION}
With the rising popularity of AI applications such as ChatGPT~\cite{chatgpt-3.5}, self-driving~\cite{9007413}, and AI agent~\cite{wang2024mobileagentv2mobiledeviceoperation}, 
machine learning has become integral to both cloud and edge computing environments. 
However, the significant computational demands of these AI workloads cannot be efficiently satisfied by CPUs or even GPUs. 
\feh{To maximize the utilization of hardware computing resources, 
manufacturers have introduced specialized AI accelerators known as Neural Processing Units (NPUs). 
There are two primary types of NPU designs: inter-core connected NPUs and monolithic NPUs (lacking inter-core connections). 
In this paper, we mainly focus on the inter-core connected NPU, which is a more powerful NPU implementation that adopts a data flow architecture.
Examples of such NPUs include the Graphcore IPU~\cite{IPU}, AWS NeuronCore~\cite{NeuronCore}, Tenstorrent~\cite{tenstorrent}, and Groq~\cite{Groq}.  
% manufacturers have introduced specialized AI accelerators known as Inter-core Connected Neural Processing Units (NPUs). 
% While there are several NPU designs that do not incorporate inter-core connection (and we refer to these NPUs as monolithic NPUs), 
% these tend to have limited computing resources and are primarily used in mobile devices, such as Intel and Apple SoCs. 
% In contrast, the most powerful NPU implementations adopt a data-flow architecture, 
% exemplified by the Graphcore IPU~\cite{IPU}, AWS NeuronCore~\cite{NeuronCore}, Tenstorrent~\cite{tenstorrent}, and Groq~\cite{Groq}. 
These inter-core connected NPUs demonstrate superior performance by efficiently leveraging the data flow inherent in AI workloads. 
Recent studies~\cite{liu2024scaling} have shown that by utilizing inter-core connections, 
the IPU can outperform the A100 GPU by a factor of 3.3x, while using only 80\% of the available FLOPS.}

% With the rising popularity of AI applications such as ChatGPT~\cite{chatgpt-3.5}, self-driving, and artificial general intelligence (AGI), 
% machine learning has become integral to both cloud and edge computing environments. 
% However, the significant computational demands of these AI workloads cannot be efficiently satisfied by CPUs or even GPUs alone. 
% \isca{To maximize the utilization of hardware computing resources, manufacturers have introduced specialized AI accelerators that adopt a data-flow architecture. 
% These are known as Inter-core Connected Neural Processing Units (NPUs), such as the Graphcore IPU~\cite{IPU}, AWS NeuronCore~\cite{NeuronCore}, Tenstorrent~\cite{tenstorrent}, and Groq~\cite{Groq}. 
% These dedicated AI accelerators deliver superior performance by efficiently harnessing the data flow inherent in AI workloads. 
% Recent studies~\cite{liu2024scaling} have demonstrated that leveraging inter-core connection, the IPU can outperform the A100 GPU by a factor of 3.3x despite using only 80\% of the FLOPS.}

Meanwhile, current AI models also exhibit significant variation in model size, leading to diverse requirements for hardware resources.
For example, a classic ResNet-50 model contains only 25 million parameters~\cite{resnet}, 
while contemporary large language models (LLMs) like Llama3 offer a wide spectrum of sizes, including options with 8, 70 and 405 billion parameters~\cite{Llama}. 
Nevertheless, the hardware design of inter-core connected NPUs has been progressively scaled up, which mainly concentrates on how to accommodate these larger models,
but often results in suboptimal resource utilization for traditional, smaller ML models such as convolutional neural networks (CNNs).
% Furthermore, developers typically operate under the assumption that they have exclusive access to dedicated hardware resources, ensuring strong isolation from other tenants. 
%Therefore, there is a growing demand among cloud tenants for NPU virtualization that can provide strong isolation, configurable NPU capacity, and hardware agnosticism.
%\DD{Maybe we should briefly give our definition of NPU virtualization here. The following is a new version:
To bridge the gap between NPU hardware design and the diverse requirements of various ML models, 
there is a growing demand for \textbf{NPU virtualization} --- \emph{a technique that can provide multiple virtual NPUs for different users/tasks based on one physical NPU, and different virtual NPUs have strong isolation, configurable NPU capacity, and hardware agnosticism.}
%}

\feh{Designing a virtualization mechanism for inter-core connected NPUs presents several challenges due to their fundamentally different architectures compared to GPUs and monolithic NPUs.
First, prior efforts only explored virtualization for SIMT architectures (e.g., GPU virtualization solutions), 
but can not tackle \rebuttal{spatially programmed accelerators} like inter-core connected NPUs.  
\rebuttal{Spatially programmed accelerators} utilize a data flow architecture with the hardware topology, 
where each NPU core occupies a unique topological position and is interconnected with other cores. 
This setup allows for direct data transfers between NPU cores without the need for additional load/store operations. 
In contrast, SIMT accelerators employ thousands of identical hardware threads/cores within a von Neumann architecture, 
where computing tasks can be offloaded to any available threads or cores, and leverage global memory to exchange intermediate results.}

Second, previous efforts have primarily addressed memory virtualization within the classical memory hierarchy (cache and global memory) used by CPUs and GPUs. 
In contrast, inter-core connected NPUs employ an SRAM-centric memory system. 
To optimize memory bandwidth for NPU cores, there is no cache coherence between the on-chip SRAM and off-chip global memory. 
Instead, data transfer between global memory and on-chip SRAM is orchestrated through DMA operations at a coarse granularity,
which proves inadequate for the fixed-size, fine-grained page-level memory virtualization~\cite{Hyun2019NeuMMUAS,feng20124sNPU}.
Therefore, due to these hardware differences, virtualizing the NPU topology with the SRAM-centric memory system is essential for inter-core connected NPUs. 
However, these aspects remain unaddressed by existing virtualization mechanisms, 
including the GPU virtualization~\cite{Dowty2008GPUVO, Zhang2018GNETEG, Tian2014AFG, Yeh2017PagodaFG, Vijaykumar2016ZoruaAH, Suzuki2016GPUvmGV} 
and monolithic NPU virtualization~\cite{xue2023vnpu, kim2023aurora}.

\isca{This paper introduces the first comprehensive virtualization design for inter-core connected NPUs, termed \sys.
Different from traditional virtualization approaches for CPUs and GPUs, 
\sys specifically targets topology virtualization with data flow architectures. 
It incorporates three key techniques for NPU virtualization:}
\begin{itemize}[leftmargin=2.5mm,topsep=2pt]
\item \textbf{vRouter for NPU route virtualization:} 
\isca{Given that the inter-core connected NPU features a hardware topology among its cores, 
it incorporates a router for instruction dispatch and data transmission within the on-chip network.  
Consequently, a virtual NPU must also inherit a virtual topology, 
and employ a virtual router to efficiently redirect both instruction flows and data flows from the virtual NPU core to the appropriate physical NPU core.}
\item \textbf{\minor{vChunk for memory virtualization within NPU topology:}}
\isca{\feh{The inter-core connected NPU exhibits distinct memory hierarchy for its SRAM-centric memory system \minor{and NoC-based interconnection}. 
Unlike the classical memory system that relies on load/store instructions in the cache granularity, 
NPUs employ DMA operations to transfer a large chunk of model weights from backup memory to SRAM,
and further eliminate load/store requests for intermediate results using the inter-core connection. 
Consequently, page-based memory virtualization is inefficient in the NPU scenario, as a TLB miss can obstruct substantial data transfers.
\sys introduces a range-based memory virtualization that fully utilizes NPU's memory access patterns.}}
\item \textbf{Topology mapping for NPU core allocation:}
\isca{\feh{Since a virtual NPU also requires a hardware topology for NPU cores, 
this presents a new challenge in allocating NPU cores while considering resource utilization. 
To address this problem, the hypervisor can adopt various topology mapping strategies, such as exact mapping and similar topology mapping. 
\sys analyzes these different strategies and proposes a balanced solution that optimizes both NPU utilization and performance for ML tasks.}}  
\end{itemize}
% Furthermore, as for the software components, \sys extends the hypervisor to manage all hardware resources as well as the topology information of virtual NPUs. 

We have developed a prototype of \sys on both an FPGA platform (Chipyard~\cite{chipyard}) for micro tests and a software simulator (DCRA~\cite{orenesvera2024dcra, DCRA-GIT}) to evaluate large-scale real-world ML applications. 
Our implementation extends the Gemmini NPU~\cite{Gemmini} with inter-core connections, featuring an architecture akin to the Graphcore IPU~\cite{IPU}. 
On the software side, we have modified the KVM module in the Linux kernel to manage all meta tables for virtual NPUs. 
\rebuttal{Evaluation results show that the hardware extensions for NPU virtualization incur only negligible overhead (about $1\%\sim2\%$) in micro-benchmark, 
and less than \DAHU{1\%} reduction in end-to-end evaluations. 
With the support of virtual topology, \sys achieves higher resource utilization through more flexible NPU core allocation. 
As a result, in multi-task scenarios, \sys delivers up to a \DAHU{1.92x} and \DAHU{1.28x} performance improvement for Transformer and ResNet models compared to the SOTA virtualization mechanism like MIG.}

\section{BACKGROUND}
\subsection{Inter-core Connected NPU Architecture}

\begin{figure*}[htp]
    \setlength{\abovecaptionskip}{0pt}
    \setlength{\belowcaptionskip}{-10pt}
    \includegraphics[width=0.9\linewidth]{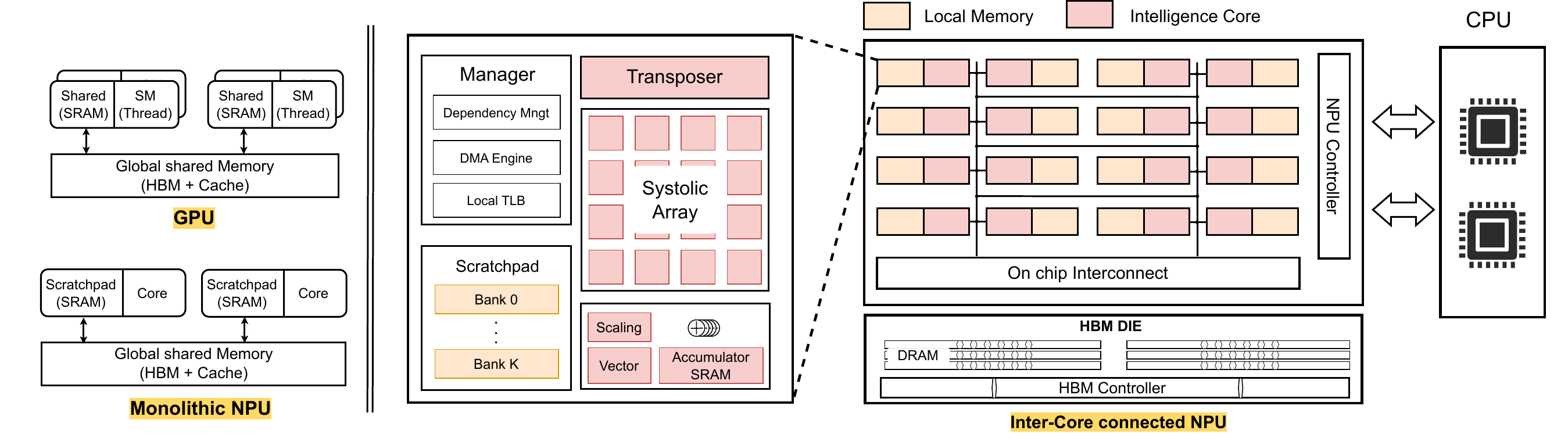}
    \caption{\textbf{An inter-core connected NPU contains multiple cores with the interconnection, on-chip SRAM and off-chip HBM.}}
    \label{fig:back:npu}
\end{figure*}

The inter-connected NPU represents an emerging class of data flow accelerators designed for AI applications. 
While there are various NPU implementations such as Graphcore IPU~\cite{IPU}, AWS's NeuronCore~\cite{NeuronCore}, Tenstorrent~\cite{tenstorrent}, 
DOJO~\cite{DOJO}, Sambanova~\cite{prabhakar2022sambanova}, Simba~\cite{shao2019simba}, MTIA~\cite{firoozshahian2023mtia}, Cerebras~\cite{lie2023cerebras} and Groq~\cite{Groq}, 
these NPUs exhibit several common characteristics. 
As depicted in Figure~\ref{fig:back:npu}, a typical inter-core connected NPU comprises multiple NPU cores, each equipped with its own on-chip interconnection and SRAM. 
To optimize memory bandwidth for AI workloads, these NPUs leverage (1) inter-core connections to directly transfer intermediate results to target nodes, 
and (2) a simplified memory architecture utilizing large local memory (i.e., scratchpad) without address association or coherence with global memory. 
The data transfer between global memory and local memory is facilitated by the DMA engine.
As for computation, NPUs are equipped with matrix computation units (e.g., systolic arrays) and vector units. 
For instance, a typical inter-core connected NPU like Graphcore's IPU features thousands of cores on a single chip and up to 900MB of on-chip SRAM.
As for the interconnection, IPU achieves all-to-all \rebuttal{65TB/s aggregated bandwidth} for on-chip network. 
Similarly, SambaNova \cite{prabhakar2022sambanova} comprises 1,040 RDU cores with 520MB of SRAM and 64GB of HBM.

\feh{Other NPUs, such as Apple NPU, Intel NPU, and Qualcomm NPU, are integrated into mobile System-on-Chips (SoCs) and may not employ inter-core connections due to hardware resource constraints.
In this paper, we do not concentrate on these NPU architectures; instead, we can apply existing solutions~\cite{xue2023vnpu,kim2023aurora} to address the virtualization challenges with such NPUs.}

% a typical NPU core generally consists of three main components: the core manager, on-chip memory, and computing units.
% The NPU core manager performs following functions: (1) retrieves and decodes NPU instructions,
% and (2) transfers data from the global Memory (e.g., HBM) to the local scratchpad memory.
% NPUs also feature a specialized memory structure and computing units.
% To maximize memory bandwidth for AI workloads, NPUs utilize a simple memory structure (scratchpad) that lacks address association and coherence, requiring software management.
% In addition, NPUs use HBM as external memory and execute DMA operations between the scratchpad and HBM.
% Regarding computing units, NPUs are equipped with multiple matrix calculation units (e.g., systolic arrays) and vector units.
% which enhance performance for matrix multiplication and vector operations.

% However, a single NPU core is often insufficient to handle computational demands of large ML models such as LLMs.
% NPUs employ network-on-chip (NoC) or inter-chip connections (ICI) to coordinate multiple NPU cores efficiently.
% Since ML workloads typically feature well-defined data flows within the computing graph,
% NPUs can utilize direct message transfers between cores to minimize memory load and store operations.
% For example, Google's TPUv4 integrates 64 cores within a single cube,
% while Graphcore's IPU includes thousands of cores on a single chip.
% These multiple cores can be partitioned into different pipeline stages to optimize the processing of AI workloads.

\subsection{Underutilization of NPU Resources}
\begin{figure}[htp]
    \setlength{\belowcaptionskip}{-5pt}
    \setlength{\abovecaptionskip}{-1pt}
    \begin{minipage}[t]{0.49\linewidth}
      \centering 
      \includegraphics[scale=0.34]{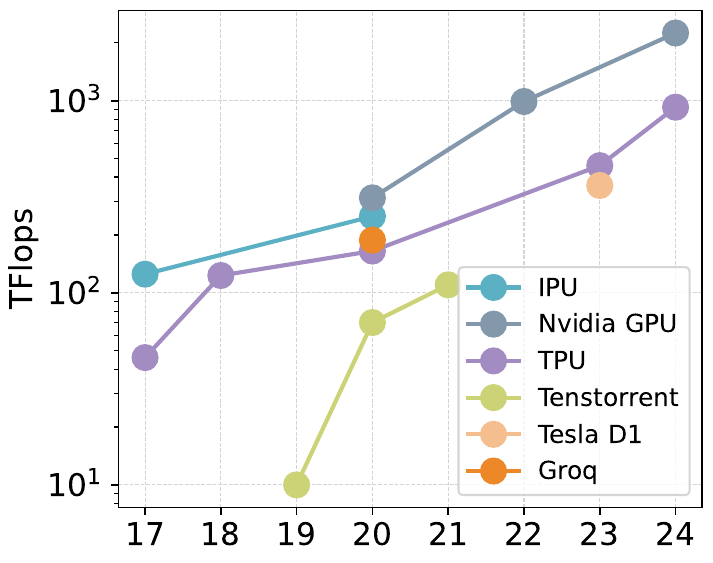}
      \footnotesize
      \textbf{(a) FLOPS}
      \end{minipage}
    \begin{minipage}[t]{0.49\linewidth}
      \centering 
      \includegraphics[scale=0.34]{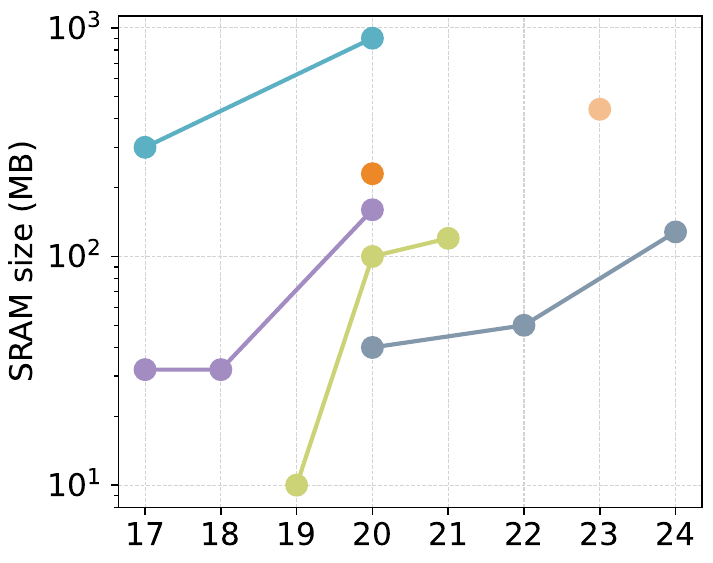}
      \footnotesize
      \textbf{(b) SRAM}
      \end{minipage}
      \caption{The evolution of NPU hardware resources: FLOPS and SRAM (2017--2024).}
      \label{fig:back:evolution} 
  \end{figure}

%\DD{Maybe this subsection should be introduced before the above ones? First explain the issues of under-utilization, and then the virtualization efforts?}
As shown in Figure~\ref{fig:back:evolution}, the latest NPUs~\cite{Groq,IPU,jouppi2017datacenter} are equipped with a large amount of computing units (>100 TFLOPS), on-chip memory (>200MB SRAM), and external high-bandwidth memory (>50GB HBM), to support the substantial computational demands of LLMs.
However, traditional small-scale ML models such as CNNs remain prevalent in both cloud and edge computing.
Platforms like Google~\cite{Googel-API} and Amazon~\cite{AWS-API} offer a range of ML APIs that cater to tasks such as label detection, object localization, and etc.
These tasks typically leverage smaller ML models, like ResNet~\cite{resnet}, GoogleNet~\cite{szegedy2014googlenet} and MXNet~\cite{chen2015mxnet}.
Moreover, the size of LLMs varies significantly, ranging from the tiny LLMs like Qwen2-0.5B~\cite{QWen2-0.5B} to giants LLMs such as GPT-3.5~\cite{chatgpt-3.5} with 175 billion parameters.
Therefore, deploying relatively small ML models on large NPU chips can lead to significant resource underutilization.

\begin{figure}[htp]
    \setlength{\abovecaptionskip}{-1pt}
    \setlength{\belowcaptionskip}{-10pt}
    \includegraphics[width=\linewidth]{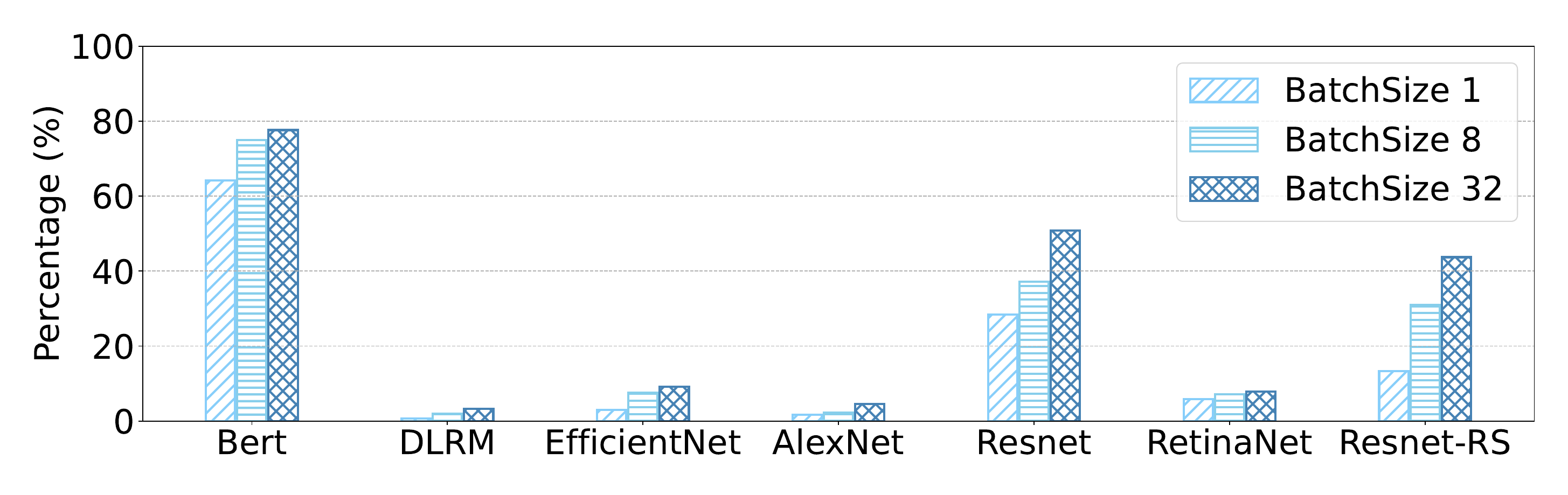}
    \caption{Overall FLOPS utilization in Google TPU with different ML workloads.}
    \label{fig:back:resource}
\end{figure}

We evaluate different ML models on a widely-used cloud NPU: Google TPU~\cite{jouppi2017datacenter}, as shown in Figure~\ref{fig:back:resource}.
The results indicate that the majority of traditional ML models utilize less than 50\% of the TPU core's FLOPS. 
Even increasing the batch size during model inference did not fully capitalize on the TPU's total FLOPS capacity. 
This underutilization stems from the imbalance between hardware resources, including FLOPS, memory size, and memory bandwidth, across various ML models.
% For instance, some ML operators, such as pooling and activation, primarily depend on the vector unit in the TPU~\cite{V10,xue2023vnpu}, 
% which may not fully leverage the throughput of the systolic array. 
Moreover, different phases during model inference exhibit distinct resource demands~\cite{patel2024splitwise}. 
The prefill phase of transformer-based models is computing-intensive, while the decode phase is memory-intensive.
\FEH{Therefore, to fully utilize all hardware resources in current large NPU chips, 
there is an imperative need for NPU virtualization to provide various resource combinations for virtual NPUs.}
% Therefore, various ML models exhibit preferences for different ratios of an NPU's hardware resources. 

\subsection{Virtualization for AI Accelerators}
\newcommand{\red}{\color[HTML]{FE0000}}%
\newcommand{\green}{\color[HTML]{008000}}%

\begin{table*}[]
    \setlength{\abovecaptionskip}{0pt} 
    \caption{\textbf{Comparison between different virtualization mechanisms of AI accelerators:} \textbf{Full or Partial Virtualization} refers to whether users are aware of operating in a virtual environment. 
    \textbf{Software Threat Model} specifies which software component is responsible for isolation between different tenants. 
    \textbf{Metric} encompasses three dimensions of resource virtualization. 
    \textbf{Number of Virtual Accelerators} indicates whether there is a constraint on the number of virtual accelerators.}
    \centering\resizebox{\textwidth}{!}{
    \label{tab:comparison}
    \begin{tabular}{c|c|c|c|ccc|c}
        \toprule
        \hline
        \multirow{2}{*}{Accelerator} & \multirow{2}{*}{Method} & \multirow{2}{*}{Full or Para Virtualization} & \multirow{2}{*}{Software Threat Model} & \multicolumn{3}{c|}{Metric}            & \multirow{2}{*}{Number of Virtual Acc.} \\ \cline{5-7}
                                     &                         &                                              &                                          & Instruction & Memory & Interconnection &                                         \\ \hline
        \midrule
        \multirow{4}{*}{GPU}         & API Forwarding~\cite{yu2020ava, wei2017vcorfu}          & \red{Para-virtualization}                          & \red{API server}                               & \green{Yes}         & \green{Yes}       & \red{No}              & \green{Unlimited}                               \\ \cline{2-8} 
                                     & MPS~\cite{mps}                     & \red{Para-virtualization}                          & \red{MPS server}                              & \green{Yes}         & \green{Yes}    & \red{No}              & \green{Unlimited}                               \\ \cline{2-8} 
                                     & MIG~\cite{mig}                     & \green{Full-virtualization}                        & \green{Hypervisor}                               & \green{Yes}         & \green{Yes}    & \red{No}           & \red{Limited, 7 in A100}                 \\ \cline{2-8} 
                                     & Time-sliced GPU~\cite{chen2017prophet, vasilas2016vgvm}         & \green{Full-virtualization}                        & \red{Scheduler}                                & \red{No}            & \red{No}       & \red{No}              & \green{Unlimited}                               \\ \hline
        \multirow{3}{*}{NPU}         & Aurora~\cite{kim2023aurora}                  & \red{Para-virtualization}                          & \red{Runtime}                                  & \green{Yes}         & \green{Yes}    & \red{No}              & \green{Unlimited}                               \\ \cline{2-8} 
                                     & V10~\cite{xue2023vnpu,V10}          & \red{Para-virtualization}                          & \green{Hypervisor}                               & \green{Yes}         & \green{Yes}    & \red{No}              & \green{Unlimited}                               \\ \cline{2-8} 
                                     & \textbf{\sys}            & \green{Full-virtualization}                        & \green{Hypervisor}                               & \green{Yes}         & \green{Yes}    & \green{Yes}           & \green{Unlimited}                               \\ \hline
        \bottomrule
    \end{tabular}}
\end{table*}
% \vspace{-5pt}

Current cloud vendors have adopted various virtualization mechanisms for AI accelerators (GPU, NPU), as shown in Table~\ref{tab:comparison}.
Considering GPU virtualization, API forwarding and time-sliced GPU represent two purely software-based virtualization mechanisms.
\textbf{API Forwarding} techniques~\cite{yu2020ava, barak2010package, duato2010efficient, giunta2010gpgpu, gupta2009gvim, gupta2011pegasus, li2011gpu, liang2011gridcuda, reano2012cu2rcu, wei2017vcorfu, xiao2012transparent, HAMI} 
can be implemented at different layers, such as rCUDA~\cite{rCUDA} and vCUDA~\cite{vCUDA} within the CUDA layer, and cGPU~\cite{cGPU} and vGPU~\cite{vGPU} at the GPU driver layer. 
Users in VMs only invoke function stubs~\cite{amiri2014paravirtualization, amiri2014rio, kuperman2016paravirtual, swift2003improving}, which are relayed by an API forwarding server to actual implementations on the host side. 
Since API forwarding does not require hardware support, it relies on the API server~\cite{duato2011enabling, kim2012snucl} to isolate computing units and memory within the GPU. 
% Additionally, some API forwarding mechanisms require modified libraries without offering full virtualization, potentially increasing development costs for users.
\textbf{Time-sliced GPU} method utilizes time-sharing mechanisms to provide multiple virtual GPUs~\cite{yu2020ava, chen2017prophet, vasilas2016vgvm, lin2016enabling}. 
Similar to API forwarding, it depends on a software module (scheduler), to isolate the time slices allocated to different VMs.
What's more, these software-based GPU virtualization mechanisms also face performance degradation due to software intervention.
Multi-process service (MPS) and multi-instance GPU (MIG) are hardware-based GPU virtualization mechanisms. 
\textbf{MPS}~\cite{mps} operates on a client-server architecture, where all tasks configured in the MPS mode dynamically send kernels to the MPS server. 
The MPS server utilizes CUDA streams to execute multiple tasks concurrently and can limit the percentage of GPU threads/SM and memory allocated to different tasks. 
However, the MPS operates in user space and is not compatible with the VM-based virtualization.
% However, the isolation of different virtual GPU tasks heavily relies on the MPS server, making it vulnerable 
% if the MPS server crashes or is compromised.
\textbf{MIG}~\cite{mig} is the latest GPU virtualization mechanism, 
which partitions a whole GPU into up to seven virtual GPUs, each of which can be passed through to a VM. 
% Unlike other GPU virtualization solutions, MIG ensures that hardware resources (e.g., memory and computing units) in the virtual GPUs are completely isolated from each other. 
However, as MIG offers the strongest security model, it sacrifices flexibility, supporting only seven virtual GPUs with several fixed configurations.

For NPU virtualization, 
recent academic research has explored para-virtualization solutions designed to monolithic NPUs. 
For instance, the Aurora~\cite{kim2023aurora} utilizes a user-space runtime to manage the mapping from virtual to physical NPU cores, 
thereby enabling swift migration between NPU tasks. 
Additionally, Xue et al.~\cite{xue2023vnpu, V10} have proposed a conceptual design for NPU virtualization with fine-grained resource sharing and isolation.
\FEH{However, these studies overlook the interconnections and topology between multiple NPU cores, which is essential for inter-core connected NPUs to optimize data flow within ML workloads.} 
% \FDH{This half sentence can be elabrated a little bit more here, since it will lead to the novelty of this work. }

\section{DESIGN OVERVIEW}
\label{s:design-overview}

% \subsection{Goal}
% \begin{itemize}[leftmargin=*,topsep=0pt]
%     \item \textbf{Full virtualization:}
%     The development of ML applications is closely related to the underlying NPU architecture. 
%     The ML compiler is required to generate the computing graph based on factors such as the total number of NPU cores, the NPU topology, and the size of the on-chip memory. 
%     Para-virtualization adds development complexity due to modifications in the NPU driver and ML compiler to accommodate different virtual NPUs.
%     In contrast, full virtualization (our goal) allows developers to use the NPU software stack without any modifications.
%     \item \textbf{High performance:} 
%     Our design aims to achieve near bare-metal performance for virtual NPUs.
%     Given the powerful computing capability of NPUs and their high-bandwidth memory, our virtualization mechanisms 
%     must be sufficiently lightweight to avoid impacting the critical path of NPU computing.
%     \item \textbf{Non-interference:}
%     \isca{In addition to resource isolation, users also require performance isolation among different tenants' virtual NPUs. 
%     Although different vNPUs have their own NPU cores, there are still some shared hardware resources within the NPU chip, 
%     such as inter-core connections and memory bandwidth for global HBM. 
%     Our design aims to achieve performance isolation, particularly focusing on inter-core connection in the current NPU architecture.}
    
% \end{itemize}

% \subsection{Overview}

\subsection{Programming Models}
\label{sub:design-overview:pm}

\definecolor{codebg}{rgb}{0.95,0.95,0.95}
\definecolor{comment}{rgb}{0.5,0.5,0.5}
\definecolor{keyword}{rgb}{0.0,0.0,0.5}
\definecolor{string}{rgb}{0.0,0.5,0.0}

% \lstset{ %
% language=C++,                % choose the language of the code
% basicstyle=\footnotesize,       % the size of the fonts that are used for the code
% numbers=left,                   % where to put the line-numbers
% numberstyle=\footnotesize,      % the size of the fonts that are used for the line-numbers
% stepnumber=1,                   % the step between two line-numbers. If it is 1 each line will be numbered
% numbersep=5pt,                  % how far the line-numbers are from the code
% backgroundcolor=\color{white},  % choose the background color. You must add \usepackage{color}
% showspaces=false,               % show spaces adding particular underscores
% showstringspaces=false,         % underline spaces within strings
% showtabs=false,                 % show tabs within strings adding particular underscores
% frame=single,           % adds a frame around the code
% tabsize=2,          % sets default tabsize to 2 spaces
% captionpos=b,           % sets the caption-position to bottom
% breaklines=true,        % sets automatic line breaking
% breakatwhitespace=false,    % sets if automatic breaks should only happen at whitespace
% escapeinside={\%*}{*)}          % if you want to add a comment within your code
% }

\lstset{
    language=C++,
    backgroundcolor=\color{white},
    basicstyle=\ttfamily\footnotesize,
    keywordstyle=\color{keyword}\bfseries,
    commentstyle=\color{comment}\itshape,
    stringstyle=\color{string},
    numbers=left,
    numberstyle=\tiny,
    stepnumber=1,
    numbersep=5pt,
    tabsize=4,
    showspaces=false,
    showstringspaces=false,
    frame=single,
    captionpos=b,
    breaklines=true,
    breakatwhitespace=true
}

\begin{lstlisting}[basicstyle=\fontsize{6.5pt}{8pt}\selectfont\ttfamily, caption={\rebuttal{Official code example of IPU programs ~\cite{IPU-PM}.}}, label={lst:ipu-code}]
Tensor v1 = graph.addVariable(FLOAT, {4}, "v1");
Tensor v2 = graph.addVariable(FLOAT, {4}, "v2");
...
Sequence prog;
prog.add(Copy(c1, v1));
...
// Create a compute set and add its execution to the program
ComputeSet computeSet = graph.addComputeSet("computeSet");
for (unsigned i = 0; i < numTiles; ++i) {
  VertexRef vtx = graph.addVertex(computeSet, "SumVertex");
  graph.connect(vtx["in"], v1.slice(i, 4));
  graph.connect(vtx["out"], v2[i]);
  graph.setTileMapping(vtx, i);
  graph.setPerfEstimate(vtx, 20);
}

prog.add(Execute(computeSet));
\end{lstlisting}
% \vspace{-10pt}

\feh{We first illustrate the programming model of the inter-core connected NPU, using the IPU as an example. 
Listing~\ref{lst:ipu-code} provides a basic demonstration of an IPU task. 
In contrast to the programming model for traditional GPU tasks, which permits offloading GPU kernel to any isomorphic hardware threads, 
\rebuttal{IPU requires developers to explicitly designate a specific NPU core for each tensor using \rebuttal{setTileMapping(tensor, coreID)}. 
The program manipulates these tensors with a set of highly parallel tasks (compute set) executed on designated NPU cores,
and fully utilizes the data flow characteristic inherent in ML tasks with inter-core communication routines. 
For instance, developers can provide a computational graph of ML tasks, 
which must be mapped onto the hardware IPU cores while considering the underlying topology. 
Furthermore, the copy primitive in the IPU software framework can leverage the on-chip network (interconnection), 
which facilitates direct data transfer between the source and destination cores.
Further details regarding the programming model for the IPU can be found on its official website~\cite{IPU-PM}.}} 
% \vspace{-10pt}
\subsection{Overview for \sys Design}

We introduce \sys, a comprehensive virtualization architecture designed for \isca{inter-core connected NPUs}, such as IPU~\cite{IPU}, tenstorrent~\cite{tenstorrent}, Groq~\cite{Groq}, etc. 
These NPUs utilize the multi-core system, scratchpad-centric memory and data flow architecture to accelerate ML tasks.
\isca{\sys consists of two major hardware extensions: vRouter (for instruction and NoC) and vChunk. Additionally, it includes an enhanced hypervisor to manage all meta-tables and resources of virtual NPUs using various mapping and allocation strategies (see in \textsection\ref{sub:design:topo-map}).}

\begin{itemize}[leftmargin=*,topsep=0pt]
    \item \textbf{vRouter:} \isca{Unlike the symmetrical cores in CPUs and GPUs, the inter-core connected NPU defines a hardware topology for multiple cores. 
    During computation, data can be transferred directly among NPU cores, thereby reducing extra memory accesses. 
    \sys introduces a new module called vRouter, to virtualize the instruction flow between the CPU core and NPU cores (i.e., dispatching NPU instructions, see in \textsection\ref{sub:design:vRouter:vInst}), 
    as well as the data flow among NPU cores (i.e., NoC, see in \textsection\ref{sub:design:vRouter:vNoC}).}
    \item \textbf{vChunk:} 
    \isca{Unlike the traditional memory hierarchy (cache with global memory) employed by CPUs and GPUs, 
    the inter-core connected NPU utilizes SRAM-centric (e.g., 900MB) and HBM/DRAM-backend memory architecture, 
    without complex hardware mechanisms such as cache coherence and association. 
    Before executing an NPU task, weights are loaded from the HBM into the local SRAM in chunk granularity. 
    % No further memory access to the backend memory occurs during the NPU execution.
    \sys employs a specialized memory virtualization mechanism called vChunk (see in \textsection\ref{sub:design:vChunk}), 
    which fully leverages NPU memory access patterns.}
\end{itemize}

\section{DETAILED DESIGN}
\label{s:design}

\subsection{vRouter: Virtualization of NPU Instruction and Interconnection}
\label{sub:design:vRouter}

\feh{As introduced in \textsection\ref{sub:design-overview:pm}, 
each NPU tensor is associated with a specific NPU core ID.
Therefore, an inter-core connected NPU must dispatch instructions and data to the appropriate cores, 
which relies on two additional hardware routers: the instruction router and the network-on-chip (NoC) router.}
% the inter-core connected NPU needs to send instructions and data to the corresponding cores,
% with two additional hardware routers: the instruction router and the NoC router.}

\myparagraph{\feh{NPU instruction router:}}
\isca{Inter-core connected NPUs comprise multiple cores at distinct topological locations. 
To achieve fine-grained access control for each core, developers utilize a dedicated NPU core ID combined with each NPU instruction. 
\feh{Once an NPU instruction is offloaded to the NPU device, a router in the NPU controller dispatches the instruction to the designated NPU core.}}

\myparagraph{\feh{NoC router:}}
Current ML tasks are structured as computing graphs, allowing NPUs to effectively leverage predefined data flows to accelerate these tasks. 
An NPU can be divided into multiple pipeline stages, with each stage dedicated to processing specific layers of ML models. 
\feh{To minimize unnecessary memory loads and stores for intermediate results, NPUs employ Network-on-Chip (NoC) among NPU cores. 
Each NoC node contains a hardware router to transmit NoC packets to the next step.}
% These interconnection mechanisms offer higher data transfer bandwidth (e.g., all-to-all 8TB/s in IPU) and lower latency compared with using the global memory. 
% Although various implementations exist, they typically necessitate a predefined sink or target node, with messages being transferred one hop/node per clock cycle.
% Compared with other processing units, NPUs can fully leverage the data flow between the multiple layers in the ML tasks.

\subsubsection{Virtualization Support for NPU Instruction Router}
\label{sub:design:vRouter:vInst}

\vspace{-15pt}
\begin{figure}[htp]
    \setlength{\abovecaptionskip}{0pt}
    \setlength{\belowcaptionskip}{-10pt}
    \includegraphics[width=\linewidth]{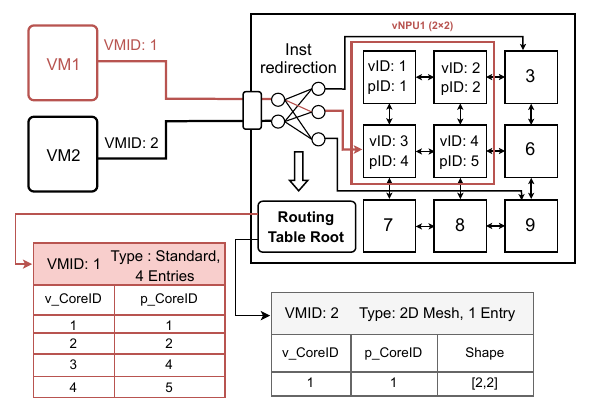}
    \caption{\textbf{Virtualization support for NPU instruction router: }\sys redirects the NPU instruction from the virtual NPU core to the physical NPU core.}
    \label{fig:design-vInstruction}
\end{figure}

\feh{Different from CPU virtualization, which requires a new mode (e.g., non-root) for virtual CPUs (temporal sharing).
The vRouter in the NPU controller focuses on redirecting NPU instructions from the virtual NPU core to the physical NPU core (spatial sharing).} 
Therefore, NPU instruction virtualization must ensure that: 
(1) Developers have the capability to access virtual NPU cores without requiring awareness of the underlying physical NPU cores; 
and (2) there is strong isolation between NPU cores across different VMs.

% Different from CPU instruction virtualization, which requires distinct privilege modes for the hypervisor and VM (temporal sharing), 
% NPU instruction virtualization concentrates on redirecting NPU instructions from the virtual NPU core to the physical NPU core (spatial sharing). 
% Given the multi-core architecture, developers must designate specific NPU cores for each pipeline stage of the ML task. 
% Consequently, NPU instruction virtualization must ensure that: 
% \FEH{(1) Developers have the capability to control virtual NPU cores without requiring awareness of the underlying physical NPU cores; 
% and (2) there is strong isolation between NPU cores across different VMs.}

\isca{\sys introduces the Routing Table (RT), a crucial data structure utilized by the vRouter. 
Similar to the page table (e.g., EPT)~\cite{barham2003xen, vt-d} used in memory virtualization, which translates virtual addresses to physical addresses, 
the routing table maps virtual NPU core IDs to physical NPU core IDs. 
Figure~\ref{fig:design-vInstruction} illustrates the various organizations of routing tables.
Each routing table is associated with a specific VM using the VMID, as well as its type and size. 
A standard routing table records each mapping between the virtual NPU core ID and its corresponding physical NPU core ID. 
However, this approach may lead to inefficient use of on-chip memory by reserving a large number of entries for the regular NPU topology.
To address this inefficiency, we propose a more specific routing table structure, 
which adopts a regular shape (e.g, 2D mesh) for NPU topology. 
This optimized structure only records the initial ID of the virtual and physical NPU core, and the shape of the virtual NPU topology.}

\isca{When receiving NPU instructions, 
the vRouter inside NPU controller translate the virtual core ID to the physical core ID according to the routing table, indexed by the VMID and the v\_CoreID. 
To alleviate potential bottlenecks in the routing table lookup, the NPU controller stores all routing tables in SRAM.}

% The routing table is compatible with different CPU-NPU interaction mechanisms. 
% There are two major types of NPU implementations:
% First, the NPU acts as a device connected via PCI-e (or other protocols). 
% Interactions between the CPU and NPU use the MMIO interface (e.g., Base Address Registers in PCI-e), 
% where each NPU core has a dedicated instruction space. 
% Here, the routing table is designed as part of the second-level page table, 
% translating the instruction space from a virtual NPU core to the corresponding physical NPU core.
% Second, the NPU acts as a co-processor and provides several specific NPU-related instructions (e.g., RoCC)~\cite{chipyard-rocc,intel-matrix,arm-matrix} for the CPU-side software. 
% In this architecture, the routing table functions as a hardware router that redirects the NPU instructions to the real physical NPU core.
% The routing table can only be managed by the hypervisor, and is transparent for the software in the VM. 
% We will introduce the routing table setup in \textsection\ref{sub:impl:hypervisor}

\vspace{-10pt}
\begin{figure}[htp]
    \setlength{\abovecaptionskip}{0pt}
    \setlength{\belowcaptionskip}{-10pt}
    \includegraphics[width=\linewidth]{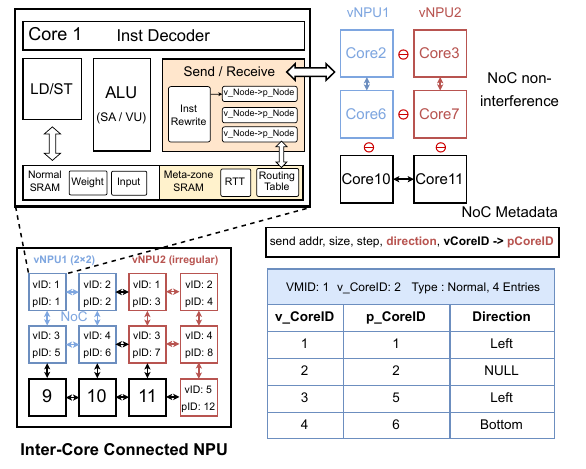}
    \caption{\textbf{NPU NoC virtualization: }\sys virtualizes the interconnection between multiple NPU cores using NoC vRouter.
	% \DD{Typo in fig: Sratchpad. Be careful about all texts in all figures...}
	}
    \label{fig:design-vRouter}
\end{figure}

\subsubsection{Virtualization Support for NoC router}

\label{sub:design:vRouter:vNoC}
% \isca{In addition to virtualizing the NPU instruction router, \sys also virtualizes the NoC router, providing a virtual topology for the virtual NPU. 
% Modern inter-core connected NPUs offer dedicated send/receive instructions, 
% enabling direct data transfer between source and destination NPU cores via the Network-on-Chip (NoC). 
% Since NPU tasks can be organized as computing graphs, inter-core communication can significantly reduce memory load/store overhead. 
% Consequently, a virtual NPU must also inherit NoC capabilities and establish inter-core connections among all virtual NPU cores.}

\isca{In addition to virtualizing the NPU instruction router, \sys also virtualizes the NoC router, providing a virtual topology for the virtual NPU. 
NoC's vRouter also relies on the routing table but requires additional information: direction, particularly for irregular NPU topologies. 
\rebuttal{For a regular NPU topology, the dimension-order routing (DOR) algorithm can be employed to effectively prevent deadlocks in the NoC network.
For example, in a 2D mesh topology, packets are routed first along the X-axis and subsequently along the Y-axis, ensuring a deadlock-free communication pattern.}
% For a regular NPU topology, the direction of NoC packets is fixed (default strategy); 
% all packets are transferred along the X-axis first, followed by the Y-axis, 
% thereby avoiding deadlock in the NoC network. 
However, when dealing with a virtual NPU with an irregular topology, 
using the default direction may result in NoC packets being transferred to an NPU node belonging to another virtual NPU.
We term this phenomenon as \textbf{NoC interference}}.
Take vNPU2 in Figure~\ref{fig:design-vRouter} as an example: 
If virtual NPU core 5 in vNPU2 wants to send packets to virtual NPU core 3, 
the default path would first transfer the packets to physical NPU core 11, and then to physical NPU core 7. 
However, this routing path passes through physical NPU core 11, which does not belong to vNPU2, 
causing NoC performance interference between different virtual NPUs.
Therefore, we provide two routing strategies for NoC virtualization:
(1) employing the default DOR strategy, which may lead to potential performance interference among virtual NPUs; 
and (2) predefining the routing direction inside the routing table to ensure that NoC packets remain confined within the virtual topology.

\isca{Figure~\ref{fig:design-vRouter} illustrates the overall design for NoC virtualization. 
We extend the original send/receive engine in each NPU core to rewrite the destination NPU core ID to the actual physical NPU core ID. 
If additional direction information is provided on the relay node, 
the vRouter will route the NoC packet based on this given direction.
Since different virtual NPU cores may have varying routing information, 
the routing table needs to be stored in the local memory (Meta-zone, see in \textsection\ref{sub:impl:arch}) of each NPU core.}

\subsection{vChunk: NPU Memory Virtualization}
\label{sub:design:vChunk}
\rebuttal{Although modern inter-core connected NPUs are equipped with substantial on-chip SRAM, 
they may also address scenarios where the model size exceeds the capacity of the SRAM and must be stored in the global HBM or DRAM. 
Therefore, \sys also supports global virtual memory for NPU cores.}
Current memory virtualization for GPUs and CPUs primarily targets the classical memory hierarchy (cache and global memory), with fine-grained load/store instructions. 
In contrast, NPUs employ a distinct memory architecture characterized by \emph{high-bandwidth, SRAM-centric local memory alongside high-capacity backend memory (HBM/DRAM)}, 
relying on coarse-grained memory access via DMA operations. 
Therefore, the traditional page-based memory virtualization may be not suitable for NPU memory virtualization 
due to its extremely high memory bandwidth with the burst phenomenon~\cite{Hyun2019NeuMMUAS}.
More specifically, to load model weights from the HBM/DRAM into the SRAM, each NPU core continuously initiates DMA operations every few cycles. 
Any TLB misses can cause a stall in numerous subsequent DMA requests, 
significantly reducing the available memory bandwidth~\cite{kim2023aurora,Hyun2019NeuMMUAS}.
\vspace{-10pt}
\begin{figure}[htp]
    \setlength{\abovecaptionskip}{0pt}
    \setlength{\belowcaptionskip}{-10pt}
    \includegraphics[width=\linewidth]{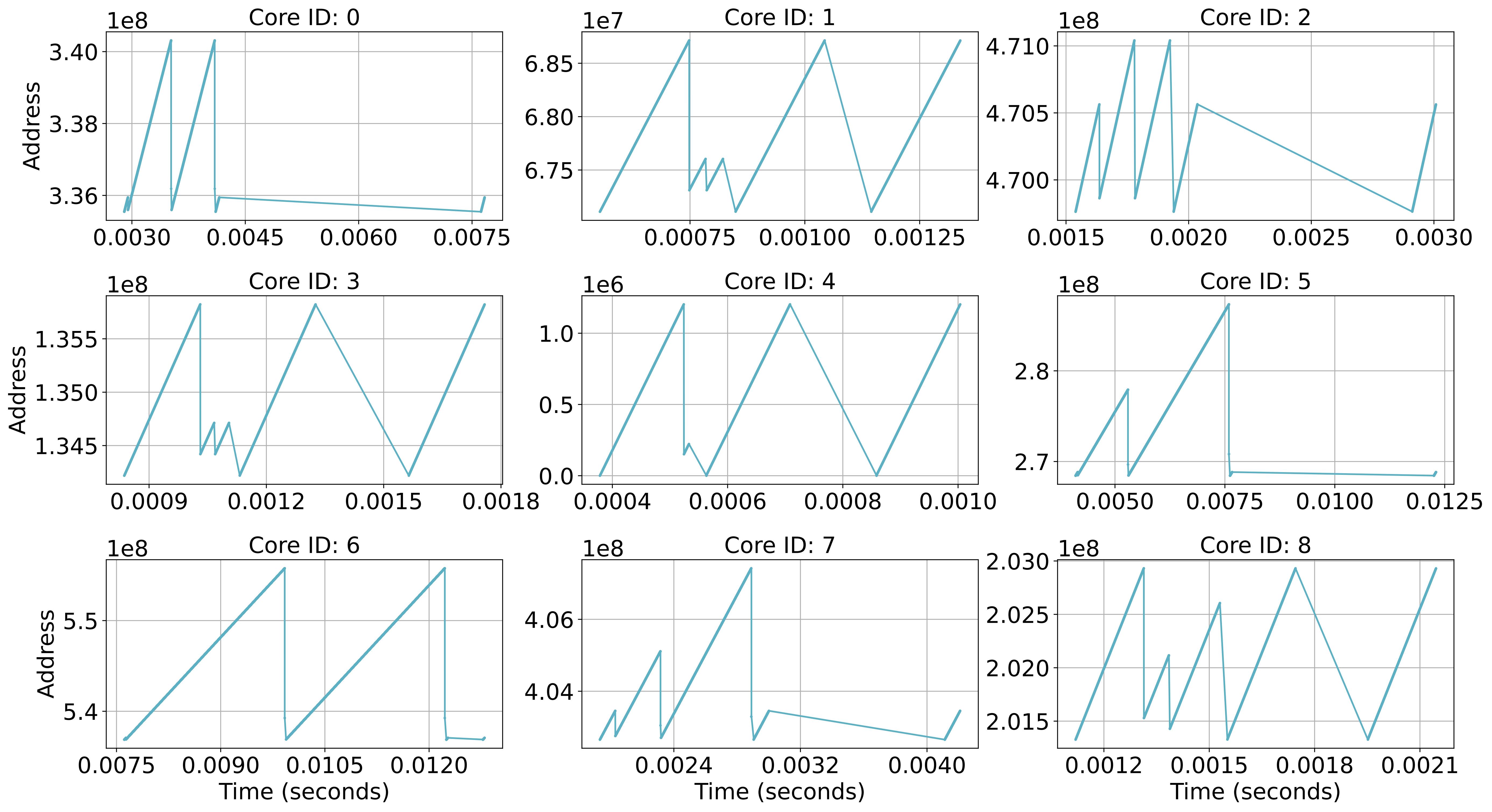}
    \caption{\rebuttal{The trace of accessed global memory addresses for the ResNet workload across different NPU cores.}}
    \label{fig:design-address}
\end{figure}

\myparagraph{Specialized memory access patterns for NPUs:}
Fortunately, the distinct memory hierarchy of NPUs with dedicated NPU workloads give rise to specialized NPU memory access patterns, 
which provides a new opportunity for NPU memory virtualization.
As for a classical NPU workflow, each NPU core first loads model weights from the global memory (HBM) into its local memory (SRAM). 
After the computation, activations or results are transferred directly via inter-core connections to the next layer,
without additional memory access to the global memory.
\rebuttal{Based on the NPU workflow, we identify three distinct memory access patterns that arise during data transfers between SRAM and HBM/DRAM.
First, model weight transfers between SRAM and HBM/DRAM predominantly occur at tensor granularity (\textbf{Pattern-1}).
Second, within a single iteration of an NPU task, 
the memory addresses of the weight tensor accessed by each NPU core typically demonstrate a monotonic increase (\textbf{Pattern-2}).
Third, given the iterative and looping nature of ML tasks, 
the tensor addresses required by each NPU core reset at the beginning of each iteration 
and repeatedly access the same set of addresses throughout iterations (\textbf{Pattern-3}).
Figure~\ref{fig:design-address} has illustrated these patterns by tracing accessed memory addresses of ResNet model across multiple NPU cores and iterations. 
In these traces, the accessed addresses demonstrate a monotonic increase within a single iteration, 
and exhibit repeated access to the same addresses across different iterations.
Moreover, the activation can be transferred directly by inter-core communication, 
thus, no additional global memory access for intermediate results during the execution.}

% Before computation, each NPU core loads model weights from the global memory (HBM) into its local memory (SRAM). 
% After computation, intermediate activations or results are transferred directly via inter-core connections. 
% Therefore, during the execution phase, no additional access to global memory is required.
% In scenarios where the entire model cannot be accommodated within the on-chip SRAM, 
% the model weight is partitioned into multiple segments, and NPU cores load different segments at different stages. 
% Based on this NPU workflow, we observe three specialized memory access patterns. 
% First, data transfer between local memory and global memory occurs only at the data chunk granularity (e.g., weight segment). 
% Second, within a single iteration, accessed memory addresses required by the NPU core typically increase monotonically. 
% Figure~\ref{fig:design-address} demonstrates the memory access trace of a classical ResNet model, 
% where required addresses of model weights increase monotonically over time, 
% while the intermediate result will be transferred directly by the inter-core connection. 
% Third, due to the repetitive nature of ML task execution, 
% the addresses of model weights remain consistent across different iterations.

\vspace{-8pt}
\begin{figure}[htp]
    \setlength{\abovecaptionskip}{-1pt}
    \setlength{\belowcaptionskip}{-13pt}
    \includegraphics[width=\linewidth]{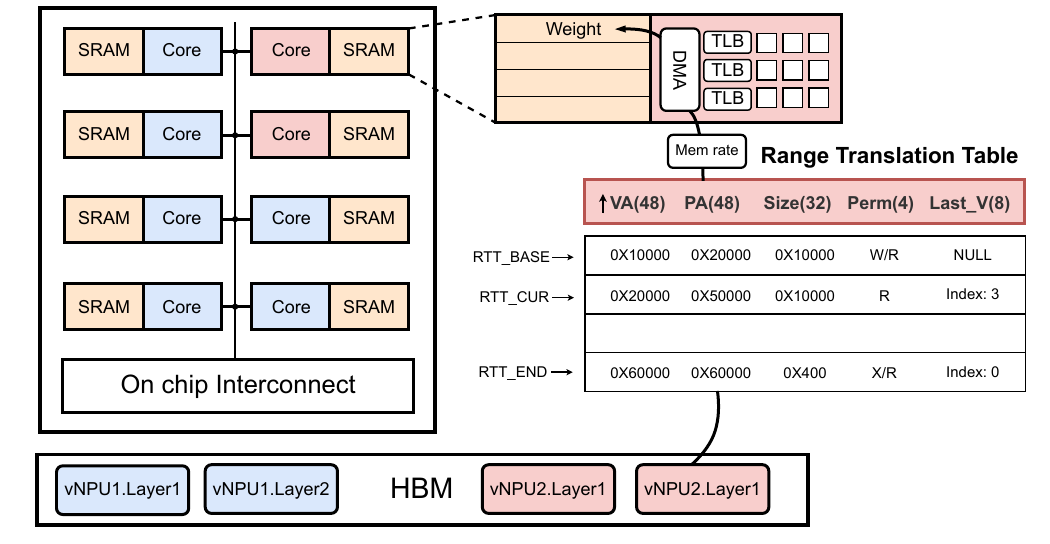}
    \caption{\sys organizes virtual memory for each NPU core at chunk granularity, and implements an efficient indexing mechanism for the range translation table.}
    \label{fig:design-vChunk}
\end{figure}

% \myparagraph{NPU memory virtualization:}
% \isca{Memory virtualization for NPUs introduces a new challenge due to the significantly higher memory bandwidth required and the memory burst phenomenon. 
% Unlike the CPU and GPU, which utilize load/store instructions, the NPU employs a DMA engine to transfer data from the global memory into the local memory of each NPU core. 
% Each cycle triggers a DMA operation, and any TLB miss can stall tens of DMA requests~\cite{Hyun2019NeuMMUAS}. 
% Current memory virtualization techniques for the CPU and GPU rely on page-based address translation~\cite{barham2003xen, vt-d} with fixed page sizes. 
% If we continue to use page-based translation for NPUs with a 4KB/2MB page size, 
% this will result in a large number of TLB misses~\cite{kim2023aurora,kim2023aurora}, thereby significantly reducing NPU memory bandwidth.}
% To address these challenges, \sys introduces a dedicated virtualization mechanism: vChunk, leveraging characteristics of ML workloads.

Therefore, \sys introduces a customized memory virtualization mechanism for NPUs called vChunk, 
which fully leverages the NPU's memory access patterns mentioned above.
vChunk replaces the \emph{fixed-page-size translation} with the \emph{range-based translation} (\textbf{Pattern-1}), 
using a novel structure called the Range Translation Table (RTT). 
Figure~\ref{fig:design-vChunk} illustrates the overall design of the RTT. 
Each entry in the table contains a virtual address (48 bits), a physical address (48 bits), the range size (32 bits) and associated permissions.
\rebuttal{Although some prior work~\cite{rangetlb,rangetranslation} has also proposed ranged translation, 
it encounters challenges in indexing entries in the range table with dynamic range sizes. 
However, the NPU workflow exhibits distinct memory access patterns, 
presenting an opportunity to optimize the indexing procedure for the range table.}  
% Unlike traditional page-based translation, where entries are easily indexed within a page table due to the fixed size, 
% indexing entries in a range-based table presents challenges due to variable range sizes. 
% To optimize the lookup process, two specific strategies are employed. 
First, \rebuttal{during the one iteration}, the model addresses required by an NPU core typically increase monotonically (\textbf{Pattern-2}), 
the next RTT entry is often the one needed. 
As a result, RTT entries are organized in ascending order of virtual addresses, 
and each NPU core maintains the index of the currently used entry (\textit{RTT\_CUR}).
\rebuttal{Second, considering the iterative/looping nature of NPU workloads, 
the model addresses required by each NPU core remain consistent across different iterations (\textbf{Pattern-3}). 
To exploit this characteristic, an additional field, \textit{last\_v}, is introduced into the RTT entry. 
This field records the index of the next entry accessed in the previous iteration. 
By leveraging this field, the accessed memory address can jump back to the initial state when transitioning to the next iteration.} 
% Second, given that model addresses required by each NPU core remain consistent iterations, 
% an additional field, \textit{last\_v}, is added into each RTT entry. 
% This field records the index of the next entry used in the previous iteration, facilitating quicker access.
During a range TLB miss, the NPU core first checks for the \textit{last\_v} value to load the corresponding range table entry into the range TLB. 
If \textit{last\_v} is either not recorded or incorrect, 
the NPU core retrieves subsequent range table entries (return to \textit{RTT\_BASE} when reaching \textit{RTT\_END}) until it retrieves the required entry. 
Finally, it updates the \textit{last\_v} field with the entry's index and adjusts the \textit{RTT\_CUR} value accordingly.

\isca{\rebuttal{In addition to memory virtualization, vChunk implements an Access Counter to locally track its memory access counts during the monitored
time window, which is similar to the prior work~\cite{kim2023aurora}.}
The NPU controller can set the maximum memory bandwidth for different virtual NPUs \rebuttal{according to user's requirements}. 
Without these memory rate restrictions, virtual NPUs may experience performance degradation due to memory interference and contention.}

\subsection{Topology Mapping Strategies for Virtual NPUs}
\label{sub:design:topo-map}

\feh{In addition to the hardware extensions required for NPU virtualization, 
the topology of the NPU introduces the challenge of core allocation during the initialization phase. 
A naive NPU core allocation strategy can lead to suboptimal resource utilization, a phenomenon we called \textbf{Topology Lock-in}. 
For instance, consider a scenario where a user requests two $3\times3$ 2D mesh virtual NPUs from a physical NPU chip with a $5\times5$ 2D mesh hardware topology. 
In such cases, the hypervisor might only be able to allocate a single virtual NPU, despite there being sufficient cores for both requests. 
This limitation occurs because the hypervisor cannot configure two virtual NPUs with the same topology that match the user's requirement. 
As a result, the topology lock-in problem leads to a wastage of approximately 64\% of the NPU cores (16 out of 25 in this example). 
Instead of allowing such inefficiency, \emph{can we explore strategies to utilize these idle NPU cores on a best-effort basis?}   
}

% \isca{Unlike CPU and GPU virtualization, where any CPU/GPU cores can be allocated to any virtual machines, NPU cores have specific locations within the NPU topology. 
% Therefore, we cannot arbitrarily allocate NPU cores to a user-defined virtual NPU.
% We define the procedure of allocating NPU cores to virtual NPUs as Topology Mapping. 
% Several alternative topology mapping strategies exist for virtual NPUs. 
% A straightforward strategy is exact mapping, where the topology of the virtual NPU matches the user-defined configuration. 
% For example, if a user requests a $3\times3$ 2D mesh virtual NPU, the hypervisor will traverse all idle NPU cores to find a set that exactly matches the user's requirements.
% While exact mapping achieves optimal performance, it can significantly reduce NPU utilization. 
% For instance, consider an NPU with 25 cores in a $5\times5$ 2D mesh topology, and a request for two virtual NPUs with a $3\times3$ 2D mesh topology. 
% Using the exact mapping strategy, the hypervisor can allocate only one virtual NPU, even though there are enough cores for both. 
% This limitation arises because the hypervisor cannot allocate two virtual NPUs with a topology that meets the user's requirements.}

To enhance NPU utilization, we propose a \emph{topology-mapping} strategy for NPU core allocation. 
This approach relaxes the constraints of topology lock-in for virtual NPUs, thereby maximizing resource utilization. 
However, the topology-mapping strategy must satisfy three requirements:
First, the number of nodes in the mapped topology must be identical to that of the original topology (\textbf{R-1}). 
Second, the mapped topology should closely resemble the original topology (\textbf{R-2}). 
Third, to ensure non-interference between virtual NPUs, the topology must remain connected (\textbf{R-3}).
% We employ the minimum topology edit distance algorithm to identify the most similar topology (the first requirement). 
% While, the second requirement (connected topology), guarantees that there is no data flow interference between virtual NPUs.

\begin{figure}[htp]
    \setlength{\abovecaptionskip}{0pt}
    \setlength{\belowcaptionskip}{-10pt}
    \includegraphics[width=\linewidth]{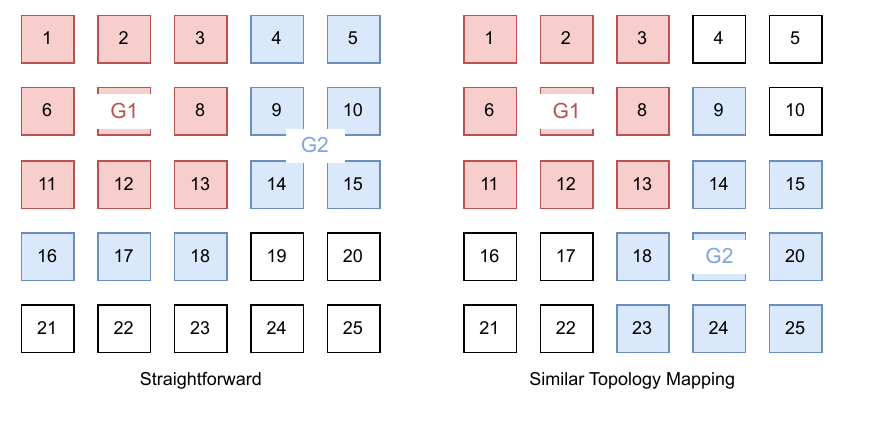}
    \caption{\textbf{Different topology mapping strategies for vNPU core allocation.} Assume the NPU consists of 25 cores arranged in a $5\times5$ 2D mesh topology. A user wants to allocate 2 vNPUs, each with a $3\times3$ 2D mesh topology.}
    \label{fig:design-topo-alloc}
\end{figure}

\begin{figure}[htp]
    \includegraphics[width=\linewidth]{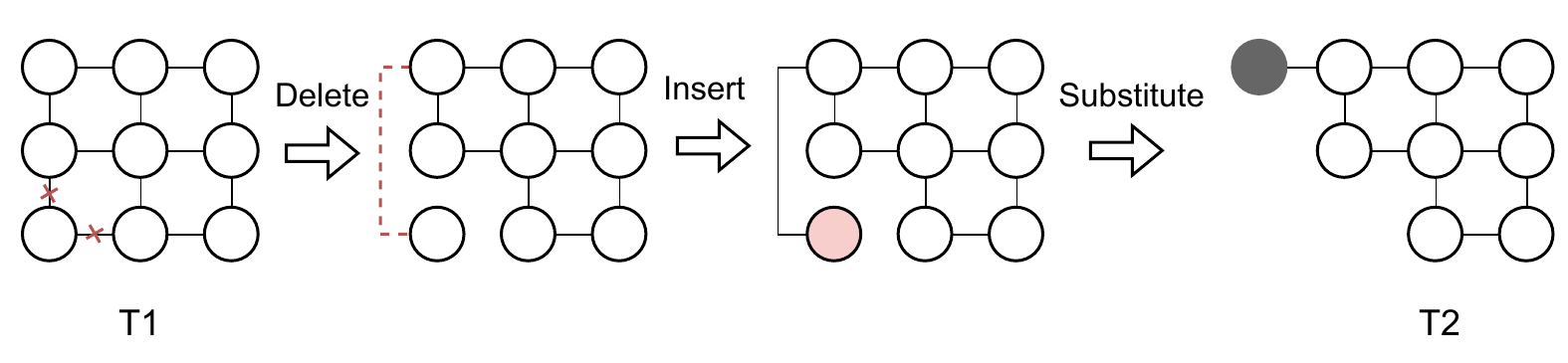}
    \setlength{\abovecaptionskip}{-15pt}
    \setlength{\belowcaptionskip}{-10pt}
    \caption{The topology edit distance between T1 and T2 is four, involving two edge deletions, one edge insertion, and one node substitution.}
    \label{fig:design-ted}
\end{figure}

\feh{Figure~\ref{fig:design-topo-alloc} illustrates NPU core allocation using different hardware topology mapping strategies. 
A straightforward approach involves allocating NPU cores based on their core IDs. 
However, it potentially leads to suboptimal performance due to increased inter-core connection overhead. 
To address this, \sys uses a minimum topology edit distance algorithm to identify a topology for the virtual NPU that is similar to the original topology.}
The topology (or graph) edit distance~\cite{hart1968formal, riesen2007speeding, riesen2009approximate, neuhaus2006fast} quantifies the dissimilarity between two topologies by determining the minimum number of edit operations required to transform one topology into the other. 
These edit operations include the insertion, deletion, and substitution of nodes and edges. 
Figure~\ref{fig:design-ted} illustrates an example: transforming T1 into T2 requires four edit operations, 
specifically two edge deletions, one edge insertion, and one node substitution.
Therefore, the topology edit distance between T1 and T2 is 4.

The pseudo-code implementation of the algorithm is shown in Algorithm~\ref{alg:ted}. 
Here, $T$ represents the entire NPU topology in hardware, 
$T'$ refers to the topology of NPU cores that are already allocated to other virtual NPUs, 
and $T\_req$ represents the topology required by the current virtual NPU. 
% Initially, all candidate NPU topologies are selected  NPU cores. 
We first select all candidate NPU topologies with required NPU cores (\textbf{R-1}).
Given that the problem of determining the minimum topology edit distance is NP-hard, 
it is essential to prune candidate topologies to minimize unnecessary computations. 
We employ three intuitive pruning strategies:
First, if a virtual NPU requires a guarantee of non-interference, 
the selected topology must be connected (Line 25), according to the \textbf{R-3} mentioned above.
% If a topology is not connected, it suggests that virtual NPU cores might interfere with other virtual NPUs during inter-core communication, leading to performance interference.
Second, for the same topology, we retain only one instance to reduce redundant calculations (Line 25).
Third, if a candidate topology matches the required topology exactly, 
we directly return this topology with the corresponding NPU cores (Line 22).
After selecting all candidate topologies, 
the topology edit distance is calculated between each candidate topology and the required topology (Line 13), 
and returns the similar topology with the minimum topology edit distance (\textbf{R-2}). 

\myparagraph{\rebuttal{Heterogeneous topology mapping:}} \rebuttal{Furthermore, the topology mapping algorithm also supports unbalanced networks and heterogeneous nodes within the topology. 
For instance, when dealing with non-uniform traffic patterns such as all-reduce, 
certain edges become more critical than others. 
In these cases, we can define a customized edge-match function (Line 8) that imposes a greater penalty if the critical path is absent in the target topology. 
More specifically, any delete or substitute operations on these critical paths will increase the edit distance.
Similarly, when accounting for the heterogeneous nodes in the NPU topology, 
additional penalties can be incorporated into the customized node-matching function (Line 3). 
For instance, if a node in the required topology is positioned close to the memory interface but is far from it in the actual mapped topology,
users can define a customized node-matching function to assign additional penalty to these nodes.
In our experience, this penalty value is determined by the difference in distances to the memory interface.
}

\myparagraph{\rebuttal{Topology fragmentation:}} \rebuttal{Even with the use of virtual topology and the similar topology mapping algorithm, 
the fragmentation problem persists during NPU core allocation. 
Since users' allocation requests cannot be anticipated in advance, 
we are unable to transform this allocation problem into a combinational problem. 
If we relax the constraints to permit a disconnected virtual topology of the NPUs, 
fragmented NPU cores can also be utilized by virtual NPUs to improve the resource utilization. 
However, this approach may introduce additional overhead associated with the conflict in inter-core communication, 
resulting in a trade-off between performance and resource utilization.
% We intend to solve this problem in our future work.
}
% Therefore, we just adopt existing solutions like tightest allocation strategy~\cite{bani2007efficient, bani2013submesh} along with the center-first strategy~\cite{1575844} to mitigate fragmentation during the allocation.
% If we relax the constraints and permit performance interference between virtual NPUs, the virtual topology of the NPUs may be allowed to be disconnected, thereby mitigating the effects of fragmentation.}

% Moreover, for a particular ML model, interactions between certain NPU cores may occur more frequently. 
% In such cases, it needs to increase the edit costs of the corresponding vertices and edges when calculating the topology edit distance (Line 3,8).
% Finally, the topology with the minimum topology edit distance is selected, and the corresponding NPU cores are assigned to the virtual NPU.

\begin{algorithm}
    % \vspace{-5pt}
    \caption{Topology Mapping Algorithm}
    \label{alg:ted}
    \begin{algorithmic}[1]
    \STATE \textbf{Procedure} \textsc{NodeMatch}($N1, N2$)
        \IF {$N1[abbr] \neq N2[abbr]$}
            \RETURN $NodeCost$ \COMMENT{A penalty is applied if the attributes of the two nodes differ.}
        \ENDIF
    % \EndProcedure
    \STATE
    \STATE \textbf{Procedure} \textsc{EdgeMatch}($E1, E2$)
        \IF {$E1$ exists but $E2$ is none}
            % \COMMENT{Different edges are assigned varying penalty values based on their importance.}
            \RETURN $E1.cost$ \COMMENT{Different edges are assigned varying penalty values based on their importance.}
        \ENDIF

    \STATE
    \STATE \textbf{Procedure} \textsc{computeTed}($T\_req, candidate$)
        \STATE $subGraph \gets T.subgraph(candidate)$
        \STATE $TED \gets$ \textsc{topo\_edit\_distance}($T\_req$, $subGraph$, $NodeMatch$, $EdgeMatch$)
        \RETURN $TED$

    \STATE
    \STATE \textbf{Function} \textsc{minTopologyEditDistance}($T, T', T\_req$)
        \STATE $minCost \gets \infty$
        \STATE $remainN \gets T.nodes() - T'.nodes()$
        \STATE $candidates = []$, $topos = []$
        \STATE $totalSubTopo \gets \textsc{comb}(remainN, T\_req.nodeNum)$
        \FOR {each $nodes$ in $totalSubTopo$}
                \IF { $nodes.topo$ is equal to $T\_req$}
                    \RETURN $nodes$
                \ENDIF
                \IF {\textsc{connected}($T, nodes$) \textbf{and} $nodes.topo$ is not in $topos$} 
                    \STATE candidates.\textsc{add}($nodes$)
                    \STATE topos.\textsc{add}($nodes.topo$)
                \ENDIF
                % \IF {\textsc{connected}($T, nodes$) \textbf{and} $nodes.topo$ is not in $topos$} \COMMENT{Pruning}
                %     candidates.\textsc{add}($nodes$)
                %     topos.\textsc{add}($nodes.topo$)
                %     % \RETURN $nodes$
                % \ENDIF
        \ENDFOR
        \STATE $pool \gets \textsc{multiprocess}(CPU.count)$ \COMMENT{Parallel}
        \STATE $results \gets \textsc{pool.map}(ComputeTED, T\_req, candidates)$
        \STATE $minTED \gets \textsc{min}(results)$
        \RETURN $minTED.nodes$

    \end{algorithmic}
\end{algorithm}
\section{IMPLEMENTATION}

\begin{figure}[htp]
    \setlength{\abovecaptionskip}{-1pt}
    \setlength{\belowcaptionskip}{-15pt}
    \includegraphics[width=\linewidth]{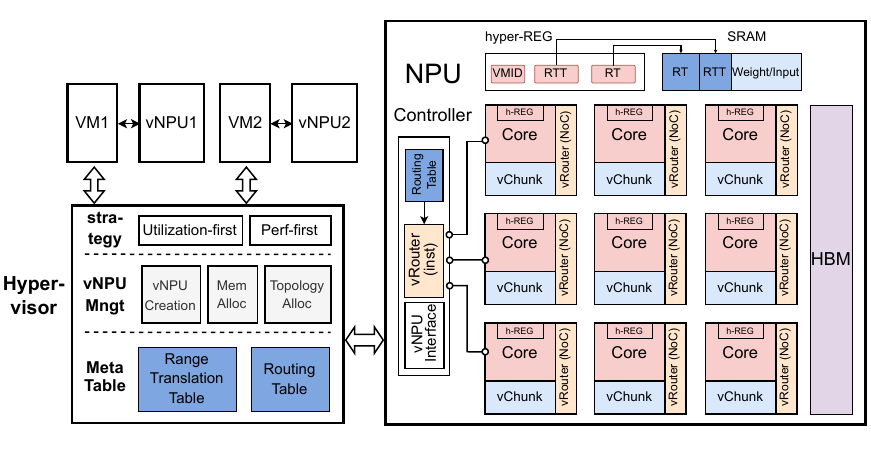}
    \caption{\textbf{The overall architecture for \sys:} \sys enhances the vRouter and vChunk modules for NPU hardware and incorporates a modified hypervisor to manage all resources of the virtual NPU.}
    \label{fig:design-overall}
\end{figure}

\subsection{Hardware Architecture for \sys}
\label{sub:impl:arch}
\isca{The right portion of Figure~\ref{fig:design-overall} illustrates the hardware architecture of \sys.
\sys first introduces a new feature: hyper mode for the NPU controller. 
Only the hyper-mode NPU controller is permitted to modify virtualization-related tables, 
such as the routing table and range translation table, and configures hyper registers (e.g., hyper-REG) for each NPU core. 
Furthermore, the hyper-mode NPU controller can collaborate with the CPU-side hypervisor.
For example, only the hypervisor is authorized to map MMIO space of hyper-mode NPU controller (e.g., PF); 
whereas guest VMs are restricted to mapping the MMIO spaces only associated with virtual NPUs (e.g., VF).}

\isca{In addition to a new mode for the NPU controller, \sys incorporates two additional hardware modules: vRouter and vChunk to achieve full virtualization. 
% The vRouter is responsible for topology virtualization, encompassing the virtualization for both instruction-flow and data-flow. 
% \sys integrates the vRouter into the NPU controller and the NoC engine, enabling redirection of NPU instructions and data flow to the appropriate NPU core. 
% The vChunk is implemented within the DMA engine to manage the loading and storing of data between the on-chip local SRAM and the global HBM.
\feh{\sys integrates the vRouter into the centralized NPU controller and the NoC engine for each NPU core. 
For NPU instruction virtualization, the vRouter within the NPU controller redirects instructions from the virtual NPU cores to the physical NPU cores based on the various types of routing tables. 
For NoC virtualization, the vRouter in each NPU core translates the destination core ID in the NoC instructions to the corresponding physical core ID, 
while further ensuring that NoC packets remain within the boundaries of the virtual topology.
The vChunk is implemented within the DMA engine to manage data movement between the on-chip local SRAM and the global HBM.
Moreover, since both the vRouter and vChunk depend on newly introduced meta tables: routing table and range translation table, 
which cannot be modified by the virtual NPU itself.
\sys partitions the on-chip SRAM into two distinct regions: the meta-zone and the weight-zone. 
The meta-zone is designated for storing all meta tables and can only be configured by the hyper-mode NPU controller. 
During execution, the NPU core is restricted from modifying the context within the meta-zone. 
To access mapping entries in these meta tables, 
the NPU core incorporates hyper-mode registers (set by the hyper-mode NPU controller), such as the base address of the routing table, etc.
As for the weight-zone, it stores the model weight and intermediate result for the NPU execution.}}

\subsection{Hypervisor Extension for \sys}

\label{sub:impl:hypervisor}
\isca{\sys requires more complicated hardware configurations for virtual NPUs, such as topology and memory. 
Therefore, the hypervisor needs to manage the internal hardware resources of each virtual NPU in a fine granularity. 
Specifically, it must organize additional meta tables, including the range translation table and routing table, 
and allocate various combinations of hardware resources to each virtual NPU in accordance with user requirements.}

\isca{First, to effectively manage the NPU's global memory (e.g., HBM) for various virtual NPUs, 
the hypervisor utilizes the traditional buddy system for memory allocation, 
and records address mappings in the range translation table. 
Unlike the page table which needs to partition blocks from the buddy system into fixed-size pages, 
\sys maps an entire block directly into the RTT entry with the block size. 
To optimize the lookup process, the hypervisor sorts RTT entries by the virtual address, 
and allocates all virtual NPU's memory during the initialization phase.
As for the virtual topology, 
% the hypervisor is responsible for configuring the routing table. 
the hypervisor adopts the similar topology mapping strategy (or exact mapping) to allocate the NPU cores for virtual NPUs.
After the allocation, the hypervisor stores the mapping between virtual core IDs and physical core IDs in the routing table. 
Furthermore, the hypervisor can also define the routing direction for each node when NoC performance isolation is required.
Both the range translation table and the routing table are managed by the CPU-side hypervisor, and finally need to be deployed on the NPU side by the hyper-mode NPU controller.}

The hypervisor maintains the abstraction of the virtual NPU, which consists of virtual NPU cores, topology, and memory. 
During the creation of a VM, the user must specify the requirements for the virtual NPU, 
including the number of NPU cores, the required topology, and the size of NPU memory. 
The hypervisor then allocates these NPU resources and configures the meta table for NPU virtualization.
NPU resources are spatially shared among different virtual NPUs.
% and the hypervisor does not need to perform a context switch (e.g., load/store the VMCS) for these virtual NPUs during VM exits.
We discuss the different sharing models of virtual NPUs in \textsection\ref{s:discussion}.  

% This distinction arises because NPU cores have the topology, and it cannot be assumed that each NPU core is isomorphic with the others.

\section{EVALUATION}
\label{s:eval}

\subsection{Experimental Setup}
\label{sub:eval:setup}
We implement a hardware prototype of vNPU on top of Chipyard~\cite{chipyard}, which is a customized SoC generator designed for evaluating full-system hardware. 
As for software components, we opt Linux-6.2 as the host kernel and Linux-6.10 as guest kernel, and modify the default KVM module in Linux kernel to manage all virtual NPU resources.
The microarchitecture of the NPU design refers to the IPU~\cite{IPU}, an inter-core connected AI accelerator.
We notice that some prior work like Aurora~\cite{kim2023aurora} and Xue et al.~\cite{xue2023vnpu} propose the para-virtualization for NPU based on the unified virtual memory (UVM), 
but does not support for the data flow architecture of inter-core connected NPUs.
We have incorporated all virtualization extensions (vChunk and vRouter) of \sys in our prototype, which can concurrently support both UVM and data flow configurations.
We evaluate the vNPU performance by running micro-tests and small ML workloads using FireSim~\cite{karandikar-firesim-isca18}, a cycle-exact, FPGA-accelerated RTL simulator,
\isca{and simulating the performance of large ML workloads with DCRA~\cite{orenesvera2024dcra, DCRA-GIT}, a distributed chiplet simulator.}
% \feh{DCRA~\cite{orenesvera2024dcra}, do we need to add a DCRA implementation here?}
% \FDH{Due to resource constraints, we implement a single-chip multi-core architecture. However, we believe the same technology can be extended to chiplet-based systems.}
The configuration is shown in Table~\ref{tab:SoC-configurations}.

\myparagraph{Comparative Systems:}
\begin{itemize}[leftmargin=*,topsep=0pt]
    \setlength\itemsep{0em}
    % \item
    % \textbf{Bare metal NPU: }
    % The NPU task runs directly on the physical NPU cores and physical memory. 
    % We utilize this configuration as the baseline for evaluating ideal NPU performance.

    \item 
    \textbf{MIG-based virtual NPU}
    \feh{Similar to the MIG in GPU virtualization, the MIG NPU offers several fixed partitions for the entire NPU chip, 
    with each partition having a predetermined sub-topology among the NPU cores. 
    Across different partitions, the MIG NPU ensures strong isolation for the NPU cores, memory, and interconnections.}

    \item
    \textbf{UVM-based virtual NPU: }
    \rebuttal{Some prior works~\cite{kim2023aurora, xue2023vnpu} have introduced NPU virtualization mechanisms that incorporate unified virtual memory but lack interconnection support. 
    Despite the architectural differences of these NPUs, 
    we still compare \sys with these works to highlight the architectural advantages of inter-core connection design for NPUs, 
    as well as to underscore the significance of topology virtualization.}
    % The NPU utilizes unified virtual memory (UVM) for data transfer between NPU cores without the support of inter-core connection. 
    % We reference prior works~\cite{kim2023aurora, xue2023vnpu} as the SOTA systems for existing NPU virtualization.

    \item
    \textbf{\sys: }
    Our solution mainly focuses on topology virtualization for inter-core connected NPUs~\cite{jouppi2017datacenter, IPU, Groq}, which is absent in the prior work.
    \sys features two additional hardware modules: vChunk and vRouter for NPU resource virtualization, and extended hypervisor for NPU's virtual resource management. 
\end{itemize}

\begin{table}[!ht] % [!ht]表格在文本中放置的位置参数（努力放在当前位置，实在放不下，将放在下一页的顶部）
\centering % 表格整体居中
\setlength{\abovecaptionskip}{0pt} 
\caption{SoC configurations used in the evaluation}
\label{tab:SoC-configurations}
\resizebox{65mm}{!}{
\begin{tabular}{|c|c|c|}
\toprule %[2pt]设置线宽     
Parameter & FPGA & SIM \\ \hline
\midrule %[2pt]  
Systolic array dimension (per tile) & 16 & 128 \\ \hline
\# of accelerator tiles             & 8 & 36 \\ \hline
Scratchpad size (per tile)          & 512KB & 30MB \\ \hline
Scratchpad size (total)             & 4MB & 1080MB \\ \hline
NoC topology                        & 2D Mesh & 2D Mesh\\ \hline
Shared L2 size (UVM)                & 2MB & / \\ \hline
Shared L2 banks (UVM)               & 8   & / \\ \hline
DRAM / HBM bandwidth                & 16GB/s & 360GB/s  \\ \hline
Tops (per tile)                     & 0.5  & 16 \\ \hline
Tops (total)                        & 4    & 576 \\ \hline
Frequency                           & 1GHz & 500MHz \\ \hline
\bottomrule %[2pt]     
\end{tabular}}
\end{table}

\subsection{Micro-test for \sys Hardware Extensions}

\subsubsection{Virtualization for Instruction Dispatch}
\begin{figure}[htp]
    \centering 
    \setlength{\belowcaptionskip}{-10pt}
    \setlength{\abovecaptionskip}{0pt}   %调整图片标题与图距离
    \includegraphics[width=3.1in]{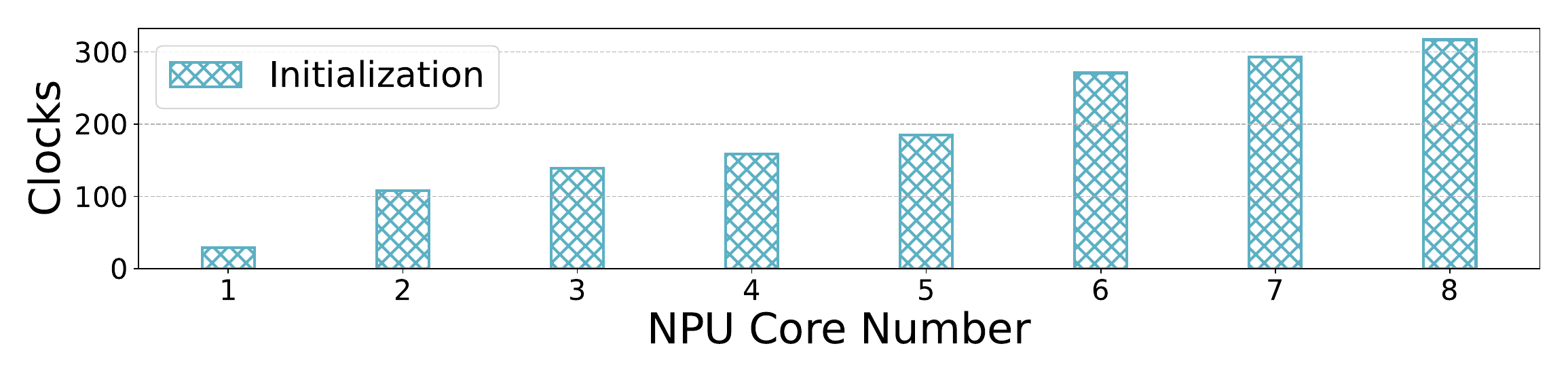}
    \caption{Configuration overhead of the routing table with different numbers of NPU cores.} 
    % \vspace{-0.1in}
    \label{fig:vInst_create}
\end{figure}

\feh{To redirect the NPU instruction from the virtual NPU core to the physical NPU core, 
we have implemented a vRouter within the NPU controller. 
We first evaluate the configuration latency for the routing table via the NPU controller,
as shown in Figure~\ref{fig:vInst_create}.}
The x-axis represents the number of NPU cores, while the y-axis (clocks) includes two overheads: querying for core availability and configuring the routing table.
The total time of routing table setup is merely a few hundred cycles, which can be neglected during the virtual NPU creation.

\begin{figure}[htp]
    \centering 
    \setlength{\belowcaptionskip}{-10pt}
    \setlength{\abovecaptionskip}{0pt}   %调整图片标题与图距离
    \includegraphics[width=3.1in]{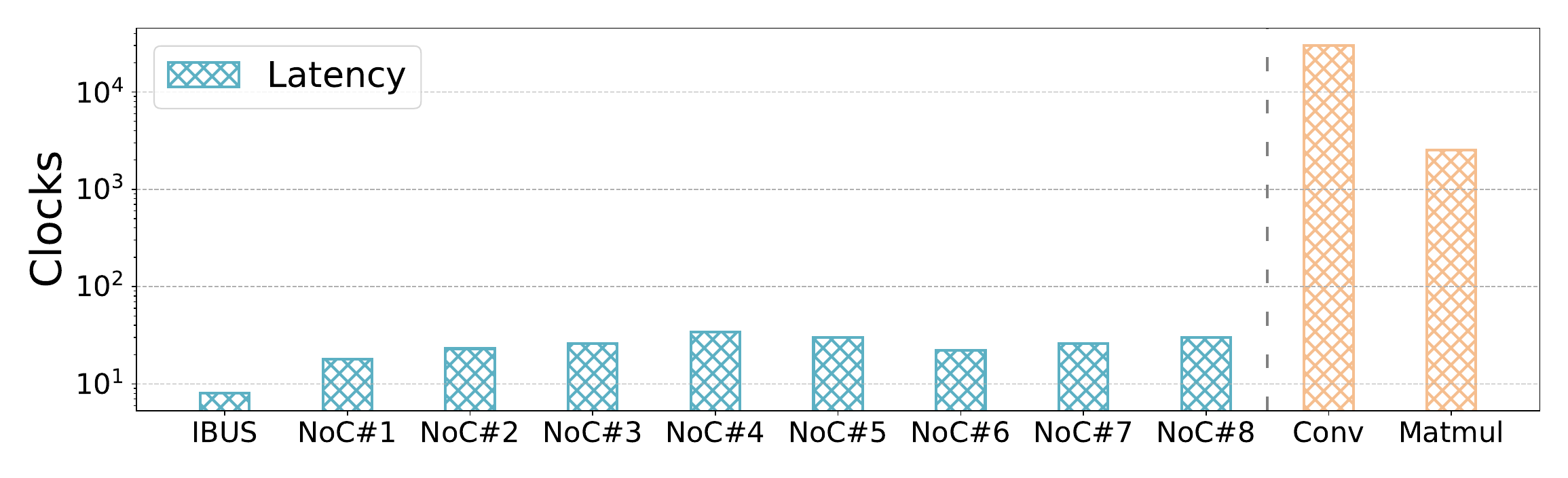}
    \caption{Latency of NPU instruction dispatch via vRouter.} 
    % \vspace{-0.1in}
    \label{fig:vInst_latency}
\end{figure}

Next, we consider the overhead of instruction dispatch during the NPU execution. 
Instructions are redirected to the corresponding NPU cores via the routing table. 
If consecutive instructions are directed to the same NPU core, 
the subsequent instructions do not need to query the routing table again.
There are two implementations for instruction routing: one using the instruction BUS 
and the other using a separate NOC to dispatch the instruction to the corresponding NPU core. 
The experimental results are shown in the left side of Figure~\ref{fig:vInst_latency}.
NoC\#1$\sim$8 indicates the delay for NPU controller dispatching instructions to core 1$\sim$8 by the separate instruction NoC.
Due to varying distances between different NPU cores in the NoC, there are slight differences in latency.
Although IBUS has the shortest and fixed latency, its transmission structure lacks scalability in multi-core systems. 
On the right side of the Figure~\ref{fig:vInst_latency},
we illustrate the execution times of two common NPU instructions: convolution and matrix multiplication.
The execution times of these NPU instructions are two to three orders of magnitude longer than the instruction routing latency. 
Therefore, the cost of instruction routing latency on overall execution is negligible.

% On the right side of the Figure 7, we present two typical
% instruction execution times: Conv represents a convolution
% instruction that fully utilizes on-chip resources, and Matmul
% represents a matrix multiplication instruction that uses par-
% tial on-chip resources. Despite the varying execution times of
% these instructions, they are two to three orders of magnitude
% longer than the instruction routing latency. Therefore, the
% cost of instruction routing latency is negligible regardless of
% using IBUS or NOC.

\subsubsection{Virtualization for inter-core connection}
\label{sub:sub:eval:micro-NoC}

% \begin{figure}[h]
%     \centering 
%     %\setlength{\abovecaptionskip}{-5pt}
%     \setlength{\belowcaptionskip}{-13pt}
%     \setlength{\abovecaptionskip}{-1pt}   %调整图片标题与图距离
%     \includegraphics[width=3.1in]{python_figs/vRouter_SR2.pdf}
%     \caption{\textbf{Micro-test of the NoC vRouter:} Compare the NoC redirection with vRouter and the standard NoC instructions.} 
%     % \vspace{-0.1in}
%     \label{fig:vRouter_SR4}
% \end{figure}

\begin{table}[!ht] % [!ht]表格在文本中放置的位置参数（努力放在当前位置，实在放不下，将放在下一页的顶部）
    \centering % 表格整体居中
    \setlength{\abovecaptionskip}{0pt} 
    \caption{\rebuttal{\textbf{Micro-test of the vRouter:} Compare the NoC virtualization with vRouter and the standard NoC instructions.}}
    \label{tab:vRouter_SR4}
    \resizebox{80mm}{!}{
    \begin{tabular}{|c|cc|cc|}
    \toprule %[2pt]设置线宽    
    \hline 
    \multirow{2}{*}{\makecell[c]{Number of routing packets \\ for data transmission}} & \multicolumn{2}{c|}{Non-Virtualization} & \multicolumn{2}{c|}{Virtualization} \\ \cline{2-5}
                                           &  Send (clk) & Receive (clk)                         & vSend (clk) & vReceive (clk) \\ \hline
    \midrule %[2pt]  
    2              & 309 & 311 & 342 & 372 \\ \hline
    10             & 1430 & 1432 & 1432 & 1492 \\ \hline
    20             & 2810 & 2818 & 2822 & 2894 \\ \hline
    30             & 4236 & 4240 & 4240 & 4308\\ \hline
    \bottomrule %[2pt]     
    \end{tabular}}
\end{table}
% \vspace{-5pt}

Besides the virtualization for instruction dispatch, 
vRouter is also adopted by the NoC engine for the virtualization of inter-core connection.
we first evaluate whether the vRouter will affect the performance of data transfers, as shown in the Table~\ref{tab:vRouter_SR4}.
$Send/Receive$ represents the data transfer between NPU cores under the non-virtualization.
While, $vSend$ /$vReceive$ means the data transfer under the vRouter mechanism. 
\rebuttal{A routing packet refers to a data block that can be transmitted by a routing instruction. In our experiment, the size of a routing packet is \DAHU{2048} bytes.}
% The x-axis represents the total number of data transmission packages (each package has 128 data lines).
Experimental result indicates that the vRouter mechanism only introduces a 1\%$\sim$2\% overhead for inter-core data transfers.

\subsubsection{Different data broadcast methods for virtual NPUs: vRouter vs. memory synchronization}
\label{sub:sub:eval:sync}
\begin{figure}[h]
    \centering 
    \setlength{\belowcaptionskip}{-10pt}
    \setlength{\abovecaptionskip}{0pt}   %调整图片标题与图距离
    \includegraphics[width=\linewidth]{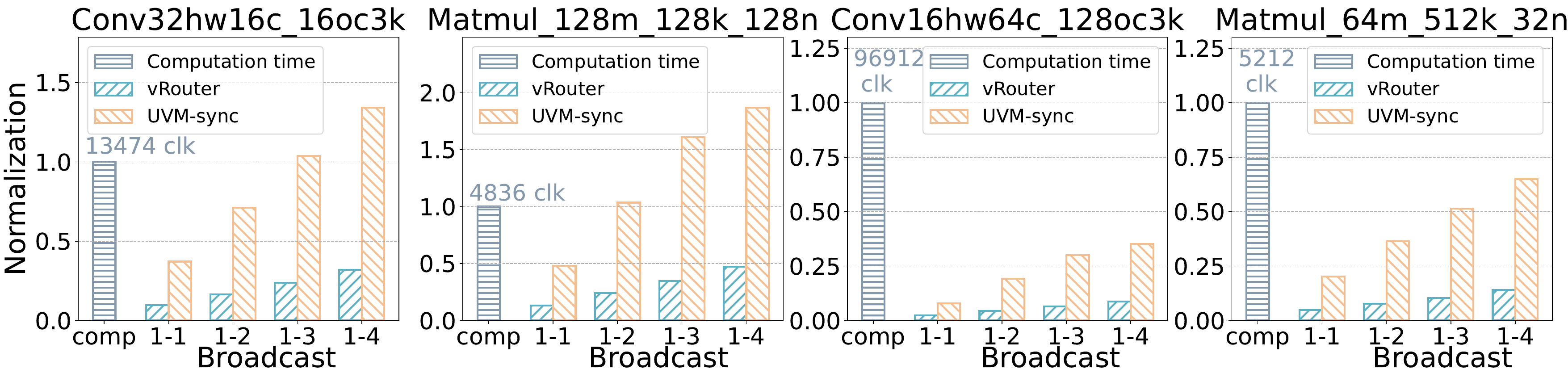}
    \caption{\rebuttal{\textbf{vRouter v.s. memory synchronization:} 
    We assess the data broadcast latency of different NPU kernels by comparing the vRouter with memory synchronization.}} 
    % \vspace{-0.1in}
    \label{fig:vRouter_multiSR}
\end{figure}

\rebuttal{Virtual NPUs can employ different data broadcast methods, 
including the vRouter mechanism proposed in this paper and synchronization through global memory. 
Many prior studies~\cite{IPU,lie2023cerebras} have highlighted the critical importance of inter-core connection for large-scale NPUs. 
In our experiments, we find that this observation also holds in the NPU virtualization scenario.
To evaluate the performance of these approaches, 
we measure the data broadcast cost across different NPU kernels with varying ratios of senders to receivers,
where each sender or receiver occupies a single NPU core.
In Figures~\ref{fig:vRouter_multiSR}, 
the label `comp' on the X-axis indicates the execution time of the NPU kernel, 
while `1:n' specifies that the kernel's execution results are broadcast to `n' nodes.
Our evaluation reveals that the vRouter mechanism achieves an average improvement of \DAHU{4.24x} compared to the global memory synchronization,
and similar experimental results also appear in the reduce operation. 
This enhancement is facilitated by vRouter's ability to exploit inter-core bandwidth and using the NoC handshake protocol for synchronization.
% This improvement arises from the vRouter's ability to exploit the additional bandwidth provided by inter-core connections. 
Furthermore, with the vRouter mechanism, the data broadcast cost is significantly lower than the NPU kernel execution time, 
allowing the broadcast overhead to be fully overlapped with kernel execution.
Conversely, when using memory synchronization, the broadcast cost for the `Matmul' kernel under the `1:4' configuration exceeds the kernel execution time. 
This imbalance can degrade the end-to-end performance of real-world NPU workloads, as discussed in \textsection\ref{sub:sub:eval:uvm}.
}

\subsubsection{Virtualization for NPU memory}

\begin{figure}[htp]
    \centering 
    \setlength{\belowcaptionskip}{-10pt}
    \setlength{\abovecaptionskip}{0pt}   %调整图片标题与图距离
    \includegraphics[width=3.1in]{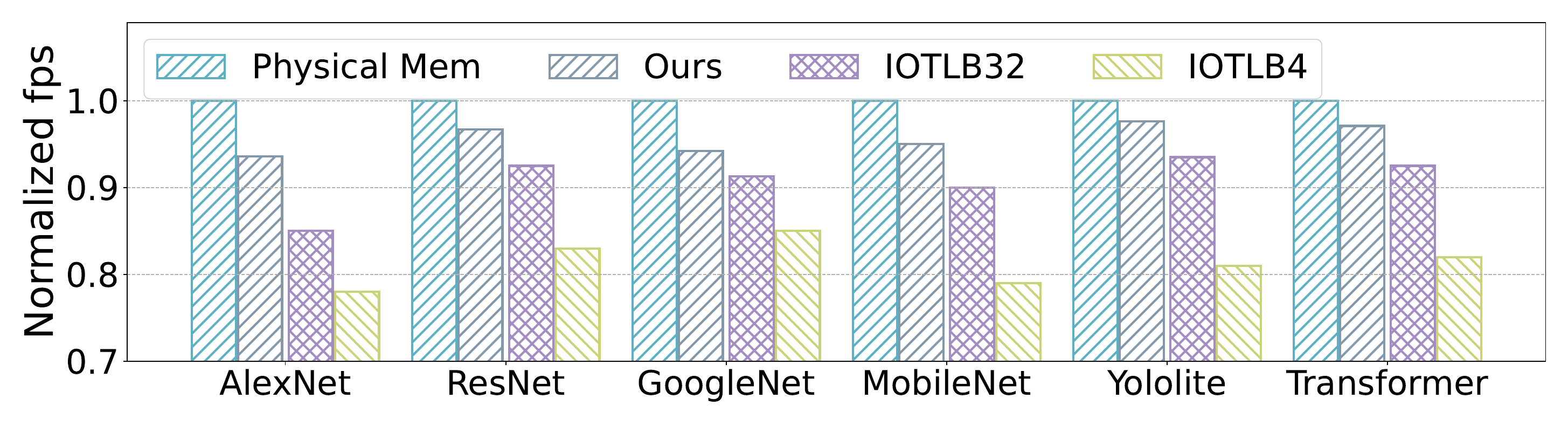}
    \caption{The normalized performance of ML workloads with different memory virtualization methods.} 
    % \vspace{-0.1in}
    \label{fig:vMem}
\end{figure}

NPU suffers from a memory burst phenomenon when loading model weights from HBM/DRAM to SRAM. 
In this process, the DMA engine within the NPU core concurrently initiates data read requests to maximize the overall memory bandwidth,
resulting a heavy translation load: triggering the translation request for every few cycles.
% Fortunately, the NPU also demonstrates a coarse-grained memory allocation along with mostly sequential memory access during the model inference. 
% This access pattern can be leveraged to minimize the translation overhead for NPU tasks

% \feh{We do not need the two-stage address translation here, similar to sNPU, we campare with the TLB-4 page table, and TLB-32 page.} 
We evaluate the translation overhead of various ML models with different configurations: physical memory, page-based and range-based translation, as shown in Figure~\ref{fig:vMem}. 
% \feh{Due to the limited SRAM in our FPGA platform, we need to partition the model to different stages.}
Compared to the ideal performance without any translation overhead (physical memory), 
the page-based translation incurs an average of \feh{20\%} overhead with only 4 TLB entries across six different models~\cite{szegedy2015going, krizhevsky2012imagenet, huang2018yolo, howard2017mobilenets, he2016deep, devlin2018bert}. 
\isca{This overhead can be partially mitigated by increasing the number of TLB entries within each NPU core; 
however, even with 32 TLB entries, the overhead remains above \feh{9.2\%}. 
vChunk employs range-level translation, and further reduce the overhead associated with range entry lookups.
\rebuttal{Using only 4 hardware range-tlb entries (144 bits for each)}, the overall overhead for vChunk is minor, remaining below \feh{4.3\% in average}.
} 
% However, enabling nested page translation for virtual NPU amplifies the translation overhead to \FEH{16\%}. 
% This increase is primarily due to a maximum of \FEH{15 times memory access} during the TLB miss in the worst-case scenario.

% \FEH{Notably, our FPGA platform only supports models with up to million parameters, 
% while the translation overhead will become more pronounced when considering larger models. 
% For instance, prior work~\cite{Hyun2019NeuMMUAS,Mosaic,kim2020batch, li2023idyll, feng20124sNPU} has demonstrated that even employing a one-stage huge-page translation results a 8\%$\sim$95\% translation overhead for GPU/NPU compared to the ideal scenario.
% In contrast, \sys adopts the range translation in the stage-two translation scheme, 
% which reduces the number of TLB misses by employing larger translation blocks. 
% Furthermore, even facing TLB misses, our RTT design only introduce one memory access at most time, 
% and achieves an average of 1.3 times memory access for one TLB miss during the entire model inference process.
% Therefore, our solution achieves the similar performance across all workloads when compared to the non-virtualization scenario that only involves one-stage translation.}

\subsection{Different Virtualization Methods for Real-world ML Applications}
\label{sub:eval:comparison}

We compare \sys to different NPU virtualization methods: UVM and MIG. 
% the SOTA NPU virtualization using unified virtual memory and SOTA GPU virtualization utilizing MIG technology. 
\rebuttal{Some previous studies~\cite{kim2023aurora,xue2023vnpu} have proposed virtual NPU abstractions, 
which are based on a unified virtual memory. 
Although it may be somewhat unfair to directly compare the performance of \sys with these solutions, 
our aim is to highlight the significance of virtual topology in the context of large-scale virtual NPUs.}
% Previous works~\cite{kim2023aurora,xue2023vnpu} have proposed virtual NPU solutions leveraging UVM. 
% However, these approaches ignore inter-core connection, relying solely on unified virtual memory for data transfer, 
% which compromise end-to-end performance due to limited memory bandwidth. 
\rebuttal{We also observe that some commercial NPUs, such as the TPU-v6e, have implemented MIG-based NPU virtualization with several fixed partitions~\cite{TPU-VM}. 
Following a similar approach, we partition the NPU topology into predefined sub-topologies and construct a comparable system, termed MIG-NPU.} 
% MIG offers several fixed configurations for virtual GPUs, and we adopt a similar strategy for NPU virtualization as another comparable method. 
% We partition the NPU topology into several fixed sub-topologies for each virtual NPU. 
In this setup, NPU cores within the same virtual NPU can establish inter-core connections, 
while the strong isolation is maintained between NPU cores located in different sub-topologies.
Due to hardware resource constraints on our FPGA platform, 
we are limited to synthesizing a maximum of 8 NPU cores and 4MB SRAM in our FPGA-based implementation. 
\rebuttal{To assess the performance of ML workloads on a larger NPU chip (e.g., 36/48 NPU cores, \DAHU{1080/1440MB} on-chip SRAM), 
which is capable of accommodating all model weights within its on-chip SRAM using the tensor partition, 
such as SOTA data-flow NPUs like Groq~\cite{Groq} and IPU~\cite{IPU}.
we modify a chiplet simulator: DCRA~\cite{DCRA-GIT}, with the \sys extension and performance profiler.}

\subsubsection{Compared \sys with the UVM-based virtual NPUs}
\label{sub:sub:eval:uvm}
\begin{figure}[htp]
    \centering 
    \setlength{\belowcaptionskip}{-5pt}
    \setlength{\abovecaptionskip}{0pt}   %调整图片标题与图距离
    \includegraphics[width=\linewidth]{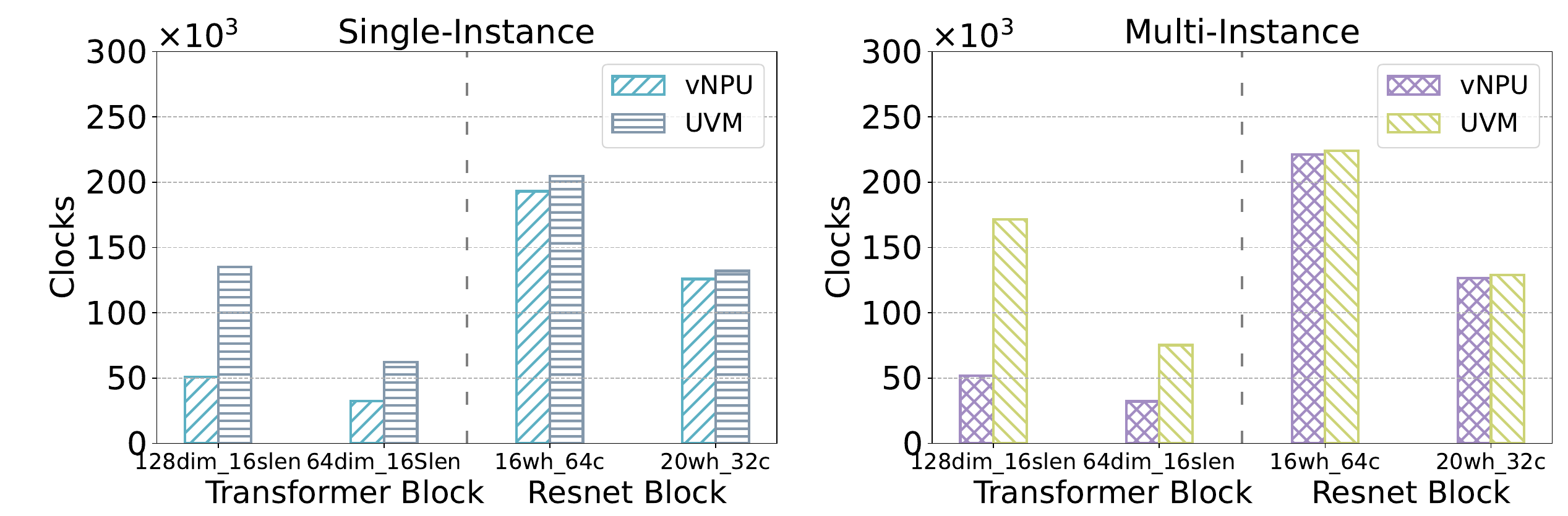}
    \caption{\rebuttal{\textbf{\sys v.s. UVM-based virtual NPU:} We evaluated the performance of different NPU virtualization mechanisms under both single-instance and multi-instance scenarios.}}
    % \vspace{-0.1in}
    \label{fig:vRouter_benchmark}
\end{figure}

% In order to test the performance of vRouter mechanism in real-world scenarios, 
% we evaluate the end-to-end performance of different ML workloads shown in Figure~\ref{fig:vRouter_benchmark}. 
% To leverage the advantages of a data-flow architecture and reduce memory access overhead, 
% we use the inter-core communication (i.e., send and receive) for inter-layer data exchange with the vRouter approach to minimize the memory wall during the NPU execution.

% Without topology virtualization, the UVM-based NPU can only leverage the unified virtual memory for data transfer.
% We compare \sys with the UVM-based virtual NPU across two classical ML workloads: ResNet and Transformer, 
% as illustrated in Figure~\ref{fig:vRouter_benchmark}. 
\rebuttal{Figure~\ref{fig:vRouter_benchmark} illustrates the performance comparison between \sys and UVM-based virtual NPUs. 
Although the primary performance improvement in \sys is driven by direct inter-core communication, which may present an inherent advantage over UVM-based NPUs,
this evaluation is intended to highlight the benefits of virtual topology routing specifically for data-flow NPUs.
In this experiment, each ML workload is executed within a dedicated virtual NPU, 
with four NPU cores allocated to each instance. 
We also assess the performance interference between multiple virtual NPUs running on a single NPU chip. 
Specifically, one virtual NPU runs the ResNet model, while another runs the Transformer model.
% The X-axis labels, Block1 and Block2, represent two network configurations with varying channel sizes (i.e., 32 for Block1 and 64 for Block2).
Our experimental results demonstrate that in a single-instance scenario, \sys achieves a \DAHU{2.29x} performance improvement for the Transformer model 
compared to the UVM-based virtual NPU. 
This improvement is primarily attributed to the virtual topology routing support provided by \sys.
However, for the ResNet model, the benefit of direct data transfer is less pronounced, yielding only a \DAHU{5.4\%} improvement in performance.
This is due to the varying layer structures in ResNet, which introduce bubbles in the data flow, ultimately leading to performance degradation.}

% In the single-task scenario, we allocate four NPU cores to execute a single ML workload. 
% % enabling a comparison of the end-to-end performance across different virtualization methods. 
% As for the multi-task test, we utilize eight NPU cores to concurrently run a ResNet task and a Transformer task, 
% with each task operating on four NPU cores. 
% The multi-task test can assess the performance isolation provided by the vRouter mechanism.
% Block1 and Block2 in the X axis represent two network configurations with different channels (i.e., 32 in Block1 and 64 in Block2).
% Our experimental results show that in vRouter architecture, the Transformer achieves a 2x performance improvement compared to UVM, 
% while the performance improvement for the ResNet is not such significant. 
% This is because the structure of each layer in the Transformer is identical, 
% allowing full utilization of pipeline parallelism. 
% In contrast, the varying structures between layers in ResNet create bubbles in the data flow, leading to performance degradation.

% Considering multi-task parallel scenarios, without the vRouter support, tasks can only run in the UVM mode, which exacerbates memory access contention between different tasks, 
Regarding performance isolation among the multi-instance scenario, 
UVM-based virtual NPUs exacerbate global memory access contention across different virtual NPUs, 
leading to an overall performance degradation of approximately 24\%. 
In contrast, \sys leverages inter-core connections, minimizing reliance on global memory. 
As a result, the performance interference between multiple tasks is negligible.

% By employing the vRouter approach, we can construct a virtual topology for each virtual NPU.
% Furthermore, vRouter also guarantees that there is no NoC interference between different virtual NPUs.
% Compared to single-task scenarios, vRouter maintains the performance of both ML tasks in multi-task scenarios, 
% whereas UVM introduces an extra overhead of 24\% for transformer blocks in average.}

% This allows NPU tasks to leverage inter-core communication capabilities without incurring any performance loss.
% Compared to the UVM, vRouter achieves an average of 35\% and 22\% performance improvement for two model configurations, respectively.}
% and mapping multiple tasks onto a single multi-core NPU without any performance loss compared to single-task scenarios.
% We unexpectedly find that in multi-task scenarios, the ResNet block's performance might surpass that in single-task scenarios slightly. 
% This could be cause by the NUMA effect, where in single-task scenarios, the ResNet block may not be allocated to the optimal cores.

\subsubsection{Compared \sys with the MIG-based Virtual NPU}
\label{sub:sub:eval:mig}
% \begin{figure}[htp]
%     \centering 
%     %\setlength{\abovecaptionskip}{-5pt}
%     \setlength{\belowcaptionskip}{-12pt}
%     \setlength{\abovecaptionskip}{-2pt}   %调整图片标题与图距离
%     \includegraphics[width=3.1in]{figs/MIG_EVAL.pdf}
%     \caption{\textbf{MIG partition v.s. \sys partition:} MIG only provides several fixed partitions, 
%     whereas \sys allows for arbitrary NPU partitions using the virtual topology.} 
%     % \vspace{-0.1in}
%     \label{fig:mig-eval}
% \end{figure}

\begin{figure}[htp]
    \centering 
    \setlength{\belowcaptionskip}{-10pt}
    \setlength{\abovecaptionskip}{0pt}   %调整图片标题与图距离
    \includegraphics[width=3.1in]{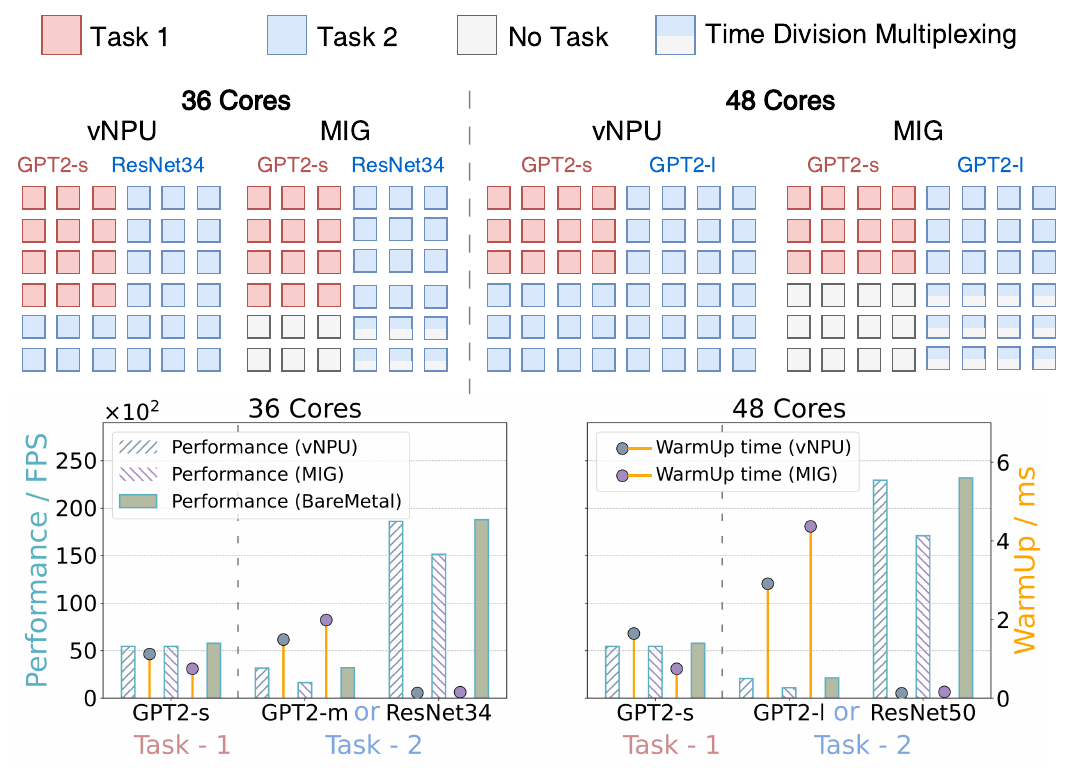}
    \caption{\rebuttal{\textbf{Performance and Warm-up latency of MIG and \sys:} We execute two virtual NPUs on a single NPU chip configured with either 36 or 48 cores. 
    In the upper half of the figure, 
    MIG employs time-division multiplexing when physical cores are less than virtual cores.
    \sys leverages a virtual topology that accommodates arbitrary topological shapes. 
    In the lower half, 
    the left Y-axis (bars) represents the performance of various NPU workloads, 
    while the right Y-axis (dots) indicates the warm-up time required before the start of NPU computation.}} 
    % \vspace{-0.1in}
    \label{fig:mig-result}
\end{figure}

\rebuttal{Current commercial NPUs, such as TPUv6-e, have introduced MIG-based virtual NPU instances with fixed topologies. 
MIG-based virtual NPUs can also leverage inter-core connections, making them a more equitable baseline when compared with \sys.}
% Moreover, we compare \sys with the MIG-based virtual NPU approach. 
% MIG has been widely adopted in the GPU virtualization, 
% and we port it into the NPU scenario as another alternative solution.
% MIG can partition an entire NPU chip into multiple sub-topologies, ensuring strong isolation between them. 
However, a significant limitation of MIG-based virtual NPU is its restriction to a limited set of fixed NPU topologies.
When user-required NPU topologies do not match the available MIG configurations, 
it can result in either under-utilization of NPU resources or the assignment of multiple virtual NPU cores to a single physical NPU core, 
as depicted in the upper half of Figure~\ref{fig:mig-result}. 
In contrast, \sys offers flexible virtual topologies for virtual NPUs, 
allowing for arbitrary configurations of NPU cores and topologies. 
% Therefore, \sys is capable of fully utilizing all NPU cores across virtual NPUs, 
% whereas MIG suffers from resource wastage due to its fixed partition.

The lower half of Figure~\ref{fig:mig-result} illustrates the task execution performance and warm-up time for both MIG-based virtualization and \sys. 
In this experiment,
%  two virtual NPUs were allocated for two distinct NPU tasks. 
% Task 1 executes on MIG1/\sys1, while Task 2 runs on MIG2/\sys2. 
we selected two classical NPU tasks for evaluation: GPT2-small/middle/large and ResNet,
and run these two NPU tasks in the two virtual NPUs with the different virtualization methods.
In our test, GPT-small consistently runs on virtual NPU 1, 
while another NPU task operates concurrently on virtual NPU 2.
\rebuttal{If the number of NPU cores is insufficient to meet user requirements (only occurs in the MIG solution), 
we temporarily share a single physical NPU core among multiple virtual NPU cores using time-division multiplexing. 
In contrast, \sys consistently allocates a sufficient number of NPU cores with fine-grained allocation for the virtual NPU; 
Although, its virtual topology may differ from the required topology.}
% \feh{The GPT models, characterized by simple and repetitive layers, 
% facilitate easier mapping of model layers to NPU cores.}
% Conversely, the ResNet model's complexity brings challenges for partitioning it among different numbers of NPU cores. 
\rebuttal{Figure~\ref{fig:mig-result} presents the evaluation results comparing the performance and warm-up time of MIG and \sys. 
The left Y-axis illustrates the concrete NPU performance, 
while the right Y-axis depicts the warmup time.}
Results indicate that for the NPU task like GPT-small, which only requires 12 NPU cores, 
the MIG-based virtual NPU in our configuration has either 18 or 24 NPU cores, wasting up to 50\% computing resources (12/24 nodes). 
In contrast, \sys can dynamically allocate the exact number (i.e., 12) of NPU cores needed by the virtual NPU, 
with the better computing resource utilization.
\rebuttal{For larger NPU workloads, such as GPT-large, which necessitates 36 NPU cores, 
\sys is capable of allocating exactly 36 NPU cores with a virtual topology. 
In contrast, a MIG-based virtual NPU can provision only up to 24 NPU cores. 
Therefore, the MIG solution must temporally share a single physical NPU core among multiple virtual NPU cores, 
resulting in up to \DAHU{1.92x} performance degradation compared to \sys.}
% While the \sys solution provides a virtual topology that includes a sufficient number of NPU cores, 
% this virtual topology may not perfectly match the required topology. 
% We analyze the performance overhead associated with this in \textsection\ref{sub:sub:eval:mapping}.}
% Notably, the performance enhancement of the GPT task \feh{is much significant in our test}. 
% Due to the simplicity of GPT's graph structure, we can efficiently map graph nodes in the GPT model to NPU cores. 
% This is because the temporal sharing used by the MIG NPU will lead to a significant number of bubbles in the execution pipeline.
\rebuttal{As for the ResNet model, the performance improvement for \sys compared with MIG NPU is minor, only \DAHU{1.28x} in average.
This is because workload imbalance in the ResNet model can be somewhat mitigated by TDM, for example, 
by binding a high-load virtual core with a low-load virtual core to share a single physical NPU core.}
% As achieving balanced workloads for ResNet is more challenging than GPT (unified decoder layers), due to its complex graph structure.
We also discuss the overhead related to the virtual topology in \textsection\ref{sub:sub:eval:mapping}

\subsubsection{Compared \sys with the bare-metal NPUs}
\label{sub:sub:eval:baremetal}
\rebuttal{In addition to comparing \sys with other NPU virtualization methods, 
we evaluate the hardware performance overhead introduced by \sys itself. 
\sys incurs only a minimal performance cost by adding a lightweight translation module that maps virtual core IDs to physical core IDs.
To quantify this overhead, we compare \sys against a bare-metal NPU configuration without any virtualization support while maintaining the same NPU topology. 
As illustrated in Figure~\ref{fig:mig-result}, 
the evaluation demonstrates that the end-to-end performance overhead introduced by \sys-based NPU virtualization is less than \DAHU{1\%} in all NPU workloads.}

\subsubsection{Warm-up time of virtual NPUs}
\label{sub:sub:eval:warmup}
\rebuttal{We also evaluate the warmup time (shown in the right Y axis in Figure~\ref{fig:mig-result}) of ML workloads under different virtualization methods. 
The warmup time is primarily spent on loading model weights from global memory into the on-chip SRAM (assume the SRAM capacity is sufficient). 
Therefore, virtual NPUs with larger memory bandwidth can achieve shorter warmup times.
In our implementation, the total memory bandwidth allocated to each virtual NPU is proportional to the number of memory interfaces associated with that virtual NPU.
However, this global memory bandwidth has minimal impact on the execution time of the NPU, 
as interactions between the on-chip SRAM and DRAM are infrequent during NPU execution.
}

\subsubsection{Different Topology Mapping Strategies}
\label{sub:sub:eval:mapping}
\begin{figure}[htp]
    \centering 
    \setlength{\belowcaptionskip}{-10pt}
    \setlength{\abovecaptionskip}{0pt}   %调整图片标题与图距离
    \includegraphics[width=3.1in]{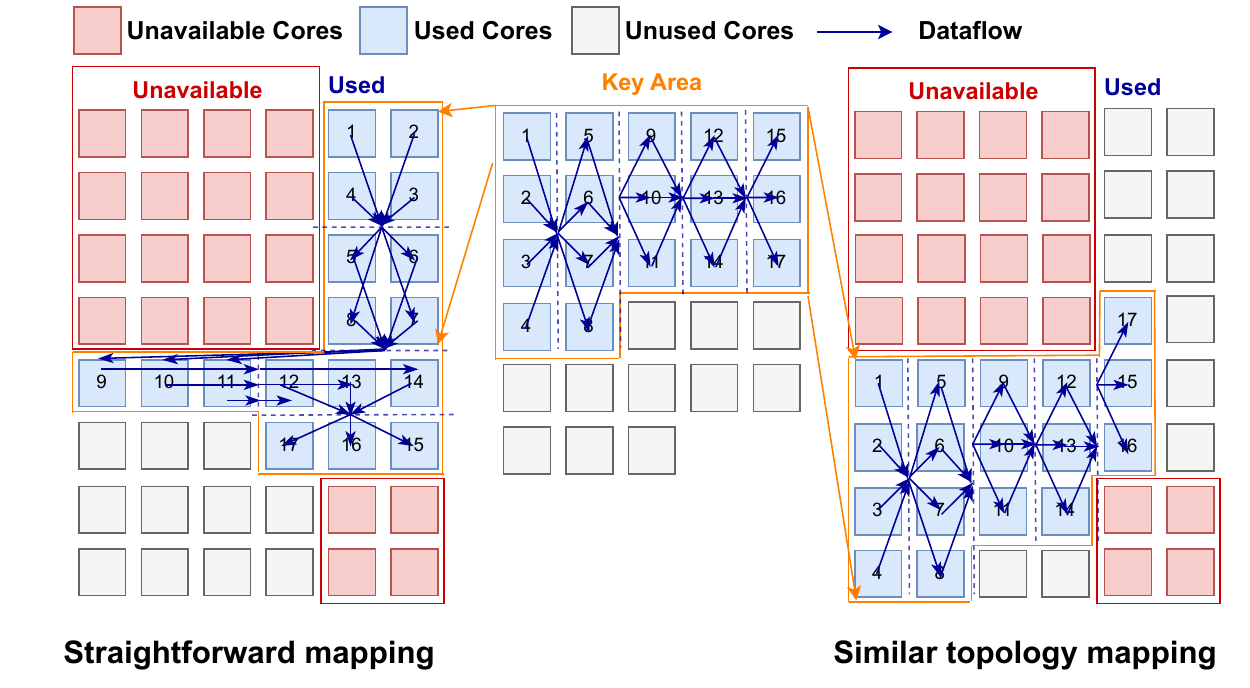}
    \caption{\textbf{Different topology mapping strategies for the ResNet model.}} 
    % \vspace{-0.1in}
    \label{fig:topo-eval}
\end{figure}

\isca{\sys adopts the topology mapping for NPU core allocation.
Figure~\ref{fig:topo-eval} illustrates different topology mapping strategies for a ResNet model.
In this scenario, the upper-left and bottom-right NPU cores have already been allocated, so the hypervisor can only assign virtual NPU cores from the remaining pool. 
We implement two topology mapping strategies: \feh{straightforward} mapping and the similar topology mapping. 
In the straightforward mapping, NPU cores are allocated based on core IDs, 
with the idle core having the smallest ID being allocated first (or zig-zag). 
In contrast, the similar topology mapping utilizes the minimum topology edit distance algorithm to allocate an NPU topology that closely aligns with the user's requirement.}

\begin{figure}[htp]
    \centering 
    \setlength{\belowcaptionskip}{-5pt}
    \setlength{\abovecaptionskip}{0pt}   %调整图片标题与图距离
    \includegraphics[width=3.1in]{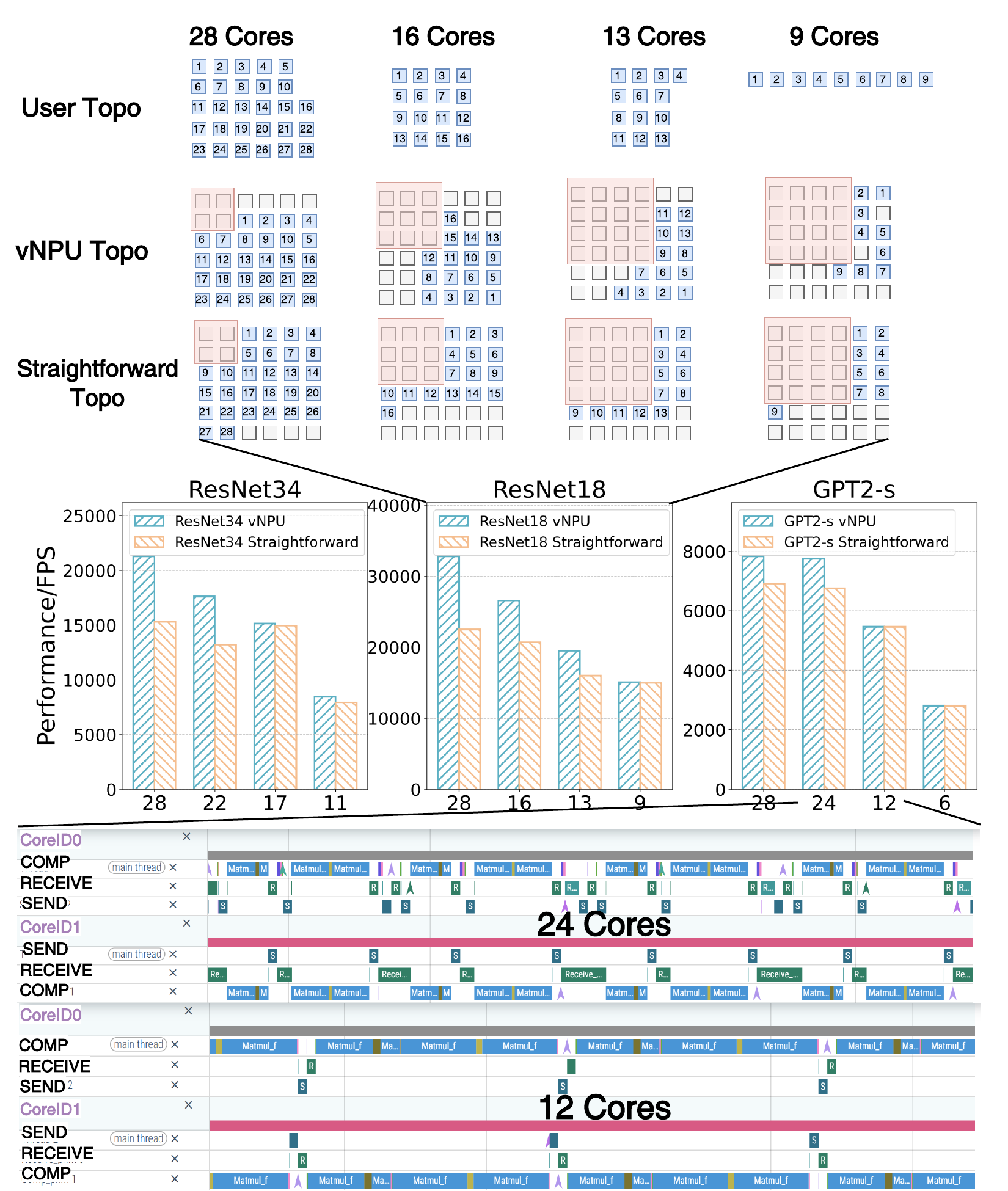}
    \caption{\rebuttal{\textbf{Performance of virtual NPU with straightforward mapping and similar topology mapping:} 
    The top part illustrates the concrete topology mappings derived from different strategies for ResNet18, 
    while the middle part depicts the performance outcomes of different ML workloads under these topology mappings,
    and the bottom part shows the core trace under different number of NPU cores with \sys topology.}} 
    % \vspace{-0.1in}
    \label{fig:topo-result}
\end{figure}

% Figure~\ref{fig:topo-result} illustrates the performance degradation observed between the straightforward mapping and similar topology mapping strategies. 
\rebuttal{In this test, we assume that the initial state of the NPU is \textbf{not empty}, 
as indicated by the red nodes in Figure~\ref{fig:topo-result}. 
We assess the performance of the virtual NPU under different topology mappings: 
the vNPU topology and a straightforward topology, and we present one of the detailed node mappings for a ResNet18 model.}
In the middle part of Figure~\ref{fig:topo-result}, 
we present the detailed performance results of various ML workloads under different topology mappings. 
The X-axis represents the number of NPU cores required by the virtual NPU, 
\rebuttal{while the Y-axis illustrates the FPS achieved by the NPU tasks.}
The evaluation results demonstrate \textbf{three trends}. 
First, the performance impact of the topology mapping strategy becomes more significant as more NPU cores are allocated. 
For example, compared with straightforward mapping strategy  (zig-zag), 
the similar topology mapping strategy achieves a 40\% improvement for ResNet34 using 28 NPU cores, but only yields a 6\% improvement with 11 NPU cores.
This is because the proportion of inter-core communication overhead decreases when multiple layers are mapped to a single NPU core.  
% \feh{Notably, the similar topology mapping strategy can identify a more optimal node mapping even when the topological shape is the same.}
Second, models with more complex graph structures are more sensitive to topology mapping strategies. 
For instance, ResNet's performance is significantly influenced by the choice of topology mapping strategy,
as \sys's mapping outperforms straightforward mapping with 42\% FPS increase in average (ResNet18/34 in 28 Cores).
In contrast, GPT models, with their layers having identical structures, 
are less sensitive to these two mapping strategies. 
Even a simple zigzag mapping can achieve 89\% performance of \sys's mapping in average.
\rebuttal{Third, regarding system scalability, while an increased number of NPU cores may introduce additional communication bubbles (bottom part of Figure~\ref{fig:topo-result}), 
the similar topology mapping strategy exhibits better scalability at the system level. 
This is because the similar topology mapping effectively reduces the overhead associated with interconnection contention, 
provided that the user-defined topology has minimal conflicts.
In contrast, a straightforward zigzag mapping, which significantly deviates from the user-defined topology, 
can lead to substantial conflicts in inter-core communication during NPU execution, 
ultimately undermining the system's scalability.
}

\subsection{Hardware Cost Analysis}
\vspace{-10pt}
\begin{figure}[htp]
    \centering 
    \setlength{\belowcaptionskip}{-10pt}
    \setlength{\abovecaptionskip}{0pt}   %调整图片标题与图距离
    \includegraphics[width=3.1in]{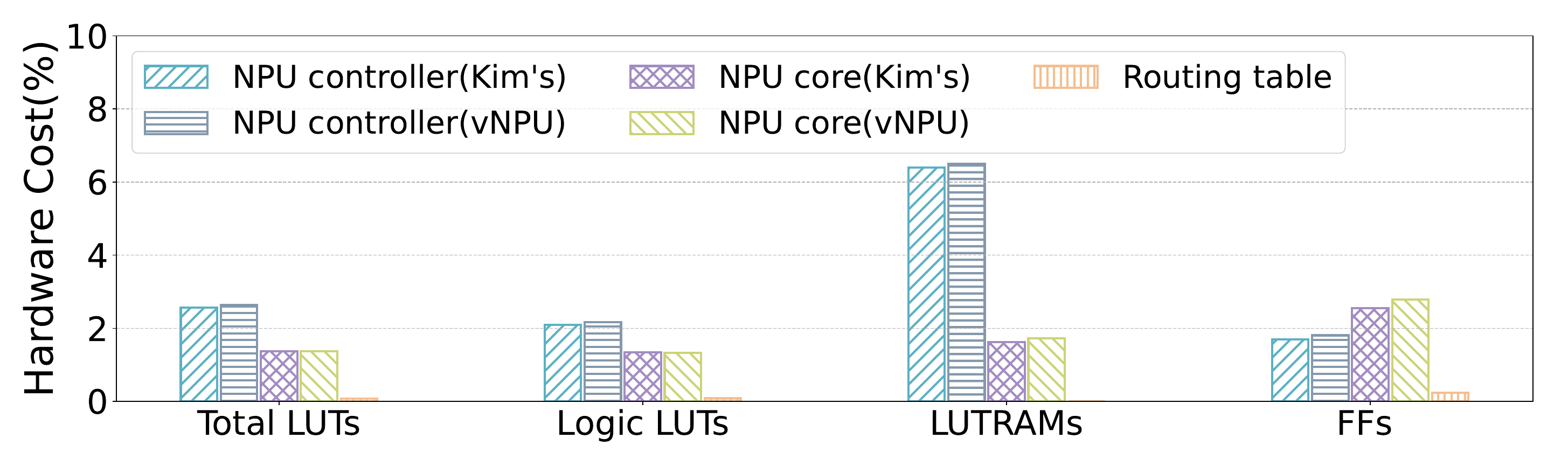}
    \caption{\textbf{Hardware resource cost:} Additional FPGA resource with different NPU virtualization mechanisms.} 
    % \vspace{-0.1in}
    \label{fig:Fpga-overhead}
\end{figure}

We synthesize the hardware resource consumption of two NPU virtualization methods on FPGA: \sys and Kim's solution \cite{kim2023aurora}, which adopts the unified virtual memory. 
Figure~\ref{fig:Fpga-overhead} illustrates the additional resources required by the vNPU on the NPU controller and cores in terms of Total LUTs, FFs, and etc. 
Different configurations are labeled as vNPU (vChunk/vRouter) and Kim's (unified virtual memory) in the legend.
Our evaluation demonstrates that the vNPU incurs minimal hardware overhead.
Both configurations require only an additional 2\% of Total LUTs and FFs. 
% Comparing vNPU and UVM configurations, the hardware overhead induced by the vNPU architecture is negligible.
For the routing table, a 128-entry configuration requires minimal FF resources for mapping storage, with LUT requirements nearly zero.

\vspace{-5pt}
\section{DISCUSSION}
\label{s:discussion}
\vspace{-5pt}
\myparagraph{Imbalanced use of VU and SA in the NPU:}
Prior work has identified imbalanced utilization of the matrix unit (e.g., systolic array, SA) and vector unit (VU) in NPU cores when executing different ML workloads. 
V10~\cite{V10,xue2023vnpu} introduces a tensor operator scheduler that coordinates diverse ML tasks within a single NPU core. 
However, this approach is not suitable for virtualization scenarios in cloud computing where different tenants do not trust each other.
To address this issue, \sys may adopt hybrid NPU cores, one optimized for matrix operations and the other for vector computations. 
Tenants can then allocate varying ratios of these two types of NPU cores according to their needs, using a virtual topology.

% \myparagraph{Security analysis:}
% \sys adheres to the same threat model as other virtualization systems, which trusts only the hypervisor (both in the CPU and NPU sides). 
% In contrast, prior work~\cite{V10,kim2023aurora} has employed a user-mode server to ensure isolation between multiple tenants; 
% however, these approaches offer a weaker isolation level compared to \sys.
% Xue et al.~\cite{xue2023vnpu} have proposed a similar solution that permits a virtual NPU to contain varying ratios of VU and SA. 
% However, their approach does not consider the topology between these NPU cores, nor does it provide a concrete implementation. 

% \myparagraph{Huge page table v.s. range translation:}
% Current AI accelerators may adopt the huge page tables or hybrid page tables~\cite{kim2020batch, lee2023memtis} to enhance address translation. 
% However, even when utilizing the huge page table, there is still a 4\% to 15\% overhead compared to the performance without TLB miss~\cite{Hyun2019NeuMMUAS, Mosaic}. 
% Furthermore, this overhead is further amplified when considering the nested page table walking in virtualization scenarios.
% \sys adopts the range translation, as there is no fragmentation for virtual NPU's memory allocation.
% Utilizing a large translation block can effectively reduce the number of TLB misses, while it becomes challenging to index the corresponding entry in the translation table. 
% Fortunately, NPUs typically exhibit sequential address access within the range granularity, and we can only retrieve the next entry in the translation table at most times.

\myparagraph{Temporal sharing v.s. spatial sharing:}
There are two methods for sharing NPU resources: temporal and spatial sharing. 
Inter-core connected NPUs typically feature multiple cores with the large on-chip memory. 
The context for NPU cores contain all model data reserved in the scratchpad, making context switching for NPUs a costly operation.
Therefore, \sys primarily utilizes spatial sharing among multiple NPU cores, without considering the expenses associated with NPU's context switch. 
However, the \sys still allows for temporal sharing if cloud vendors wish to engage in over-provisioning.

\iffalse
\myparagraph{Data transferring between CPU and NPU:}
In cloud NPUs, the NPU-side memory may be separate from the CPU-side memory. 
Therefore, data needs to be copied (using DMA~\cite{Memcpy-DMA} or NVLINK~\cite{nvlink}) from the CPU memory to the NPU memory before execution. 
There are many existing virtualization solutions like IOMMU~\cite{amd-iommu,sIOPMP} or sMMU~\cite{SMMU},
which can be leveraged by \sys to virtualize data transfers between the CPU and NPU.
Our prototype is focused on the NPU virtualization itself and is compatible with it.
\fi

\myparagraph{Address translation for graph workloads like GNN:}
\rebuttal{For graph workloads such as GNNs, 
which require large graph datasets and involve random information retrieval, 
our range-translation design may not be ideal. 
For these types of workloads, employing traditional page-level translation is recommended.}

\myparagraph{KV-cache support for NPUs:}
\rebuttal{Current commercial NPUs, such as Apple and Qualcomm NPUs, 
utilize a pre-allocated, fixed-size KV buffer. 
In our implementation, we adopt this approach as well, 
specifying a maximum size for the KV buffer in SRAM. 
The dynamic KV buffer management as well as the KV cache offloading for inter-core connected NPUs will be part of our future work.
}

% established on a unified memory system and does not incorporate this consideration.

% \vspace{-10pt}
\section{CONCLUSION}
\label{s:conclusion}
This paper presents the first comprehensive virtualization design for inter-core connected NPUs named \sys. 
\sys focuses on virtualizing specialized hardware structures for NPUs, including multi-core architecture, scratchpad-centric memory, and inter-core connection.
To achieve this, it proposes two hardware extensions: vRouter and vChunk, as well as different topology mapping strategies.
As AI models continue to evolve, incorporating more powerful NPUs, NPU virtualization will become an essential technology in future cloud computing. 
The topology virtualization presented in this paper is crucial for data flow accelerators (NPU) but is notably absent in CPU and GPU virtualization.
% We implement a prototype of \sys on an FPGA platform and achieve a 2x performance improvement compared to SOTA systems.

% First, \sys introduces vInstruction, which redirects instructions from virtual NPU cores to physical NPU cores.
% Then, \sys presents vMem, which virtualizes both on-chip scratchpad and off-chip HBM. 
% For the stage-2 translation, it utilizes a range translation, replacing the conventional page translation.
% Last, \sys incorporates vRouter, which is capable of virtualizing the interconnection between multiple NPU cores and providing a virtual topology for the virtual NPUs.
% % To better manage these hardware resources in virtual NPUs, 
% % we have also extended the existing hypervisor to configure meta-tables for \sys and allocate NPU cores with a virtual topology.
% We implement a prototype of \sys on an FPGA platform and achieve a 2x performance improvement compared to SOTA systems.

\begin{acks}
\label{sec:ack}
We sincerely thank our anonymous reviewers for their insightful suggestions,
and we also appreciate Yaodanjun Ren for the valuable advice to paper refinement.
This work is supported in part by
STI 2030—Major Projects 2021ZD0200300,
China National Natural Science Foundation (No. 62088102,623B2074),
and Huawei Central Software Institute.
Rong Zhao is the corresponding author.
\end{acks}

\end{sloppypar}

\bibliographystyle{ACM-Reference-Format}
\bibliography{references}

\end{document}